\input harvmac 
\input epsf.tex

\overfullrule=0mm

\newcount\figno
\figno=0
\def\fig#1#2#3{
\par\begingroup\parindent=0pt\leftskip=1cm\rightskip=1cm\parindent=0pt
\baselineskip=11pt
\global\advance\figno by 1
\midinsert
\epsfxsize=#3
\centerline{\epsfbox{#2}}
\vskip 12pt
{\bf Fig.\the\figno:} #1\par
\endinsert\endgroup\par
}
\def\figlabel#1{\xdef#1{\the\figno}}
\def\encadremath#1{\vbox{\hrule\hbox{\vrule\kern8pt\vbox{\kern8pt
\hbox{$\displaystyle #1$}\kern8pt}
\kern8pt\vrule}\hrule}}


\def\IR{\relax{\rm I\kern-.18em R}}
\font\cmss=cmss10 \font\cmsss=cmss10 at 7pt

\font\cmss=cmss10 \font\cmsss=cmss10 at 7pt
\def\IZ{\relax\ifmmode\mathchoice
{\hbox{\cmss Z\kern-.4em Z}}{\hbox{\cmss Z\kern-.4em Z}}
{\lower.9pt\hbox{\cmsss Z\kern-.4em Z}}
{\lower1.2pt\hbox{\cmsss Z\kern-.4em Z}}\else{\cmss Z\kern-.4em Z}\fi}
\def\IN{\relax{\rm I\kern-.18em N}}
\def\b{\circ}
\def\n{\bullet}

\def\gbbbb{\Gamma_4^{\hbox{$\scriptstyle \b \b$}\kern -8.2pt
\raise -4pt \hbox{$\scriptstyle \b \b$}}}
\def\gnnnn{\Gamma_4^{\hbox{$\scriptstyle \n \n$}\kern -8.2pt  
\raise -4pt \hbox{$\scriptstyle \n \n$}}}
\def\gnnnnnn{\Gamma_6^{\hbox{$\scriptstyle \n \n \n$}\kern -12.3pt
\raise -4pt \hbox{$\scriptstyle \n \n \n$}}}
\def\gbbbbbb{\Gamma_6^{\hbox{$\scriptstyle \b \b \b$}\kern -12.3pt
\raise -4pt \hbox{$\scriptstyle \b \b \b$}}}
\def\gbbbbc{\Gamma_{4\, c}^{\hbox{$\scriptstyle \b \b$}\kern -8.2pt
\raise -4pt \hbox{$\scriptstyle \b \b$}}}
\def\gnnnnc{\Gamma_{4\, c}^{\hbox{$\scriptstyle \n \n$}\kern -8.2pt
\raise -4pt \hbox{$\scriptstyle \n \n$}}}
\def\Rbud#1{{\cal R}_{#1}^{-\kern-1.5pt\blacktriangleright}}
\def\Rleaf#1{{\cal R}_{#1}^{-\kern-1.5pt\vartriangleright}}
\def\Rbudb#1{{\cal R}_{#1}^{\circ\kern-1.5pt-\kern-1.5pt\blacktriangleright}}
\def\Rleafb#1{{\cal R}_{#1}^{\circ\kern-1.5pt-\kern-1.5pt\vartriangleright}}
\def\Rbudn#1{{\cal R}_{#1}^{\bullet\kern-1.5pt-\kern-1.5pt\blacktriangleright}}
\def\Rleafn#1{{\cal R}_{#1}^{\bullet\kern-1.5pt-\kern-1.5pt\vartriangleright}}
\def\Wleaf#1{{\cal W}_{#1}^{-\kern-1.5pt\vartriangleright}}
\def\Cleaf{{\cal C}^{-\kern-1.5pt\vartriangleright}}
\def\Cbud{{\cal C}^{-\kern-1.5pt\blacktriangleright}}
\def\Crleaf{{\cal C}^{-\kern-1.5pt\circledR}}


\Title{\vbox{\hsize=3.truecm \hbox{SPhT/03-104}}}
{{\vbox {
\bigskip
\centerline{Statistics of planar graphs viewed from a vertex:}
\medskip
\centerline{A study via labeled trees} 
}}}
\bigskip
\centerline{J. Bouttier\foot{bouttier@spht.saclay.cea.fr}, 
P. Di Francesco\foot{philippe@spht.saclay.cea.fr} and
E. Guitter\foot{guitter@spht.saclay.cea.fr}}
\medskip
\centerline{ \it Service de Physique Th\'eorique, CEA/DSM/SPhT}
\centerline{ \it Unit\'e de recherche associ\'ee au CNRS}
\centerline{ \it CEA/Saclay}
\centerline{ \it 91191 Gif sur Yvette Cedex, France}
\bigskip
\noindent 
We study the statistics of edges and vertices in the vicinity of a
reference vertex (origin) within random planar quadrangulations and Eulerian 
triangulations. Exact generating functions are obtained for theses 
graphs with fixed numbers of edges and vertices at given geodesic 
distances from the origin. Our analysis relies on bijections with labeled trees,
in which the labels encode the information on the geodesic distance from the origin. 
In the case of infinitely large graphs,
we give in particular explicit formulas for the probabilities that the origin
have given numbers of neighboring edges and/or vertices, as well as explicit
values for the corresponding moments.

\Date{07/03}

\nref\TUT{W. Tutte, 
{\it A Census of Planar Maps}, Canad. Jour. of Math. 
{\bf 15} (1963) 249-271.}
\nref\BIPZ{E. Br\'ezin, C. Itzykson, G. Parisi and J.-B. Zuber, {\it Planar
Diagrams}, Comm. Math. Phys. {\bf 59} (1978) 35-51.}
\nref\DGZ{P. Di Francesco, P. Ginsparg
and J. Zinn--Justin, {\it 2D Gravity and Random Matrices},
Physics Reports {\bf 254} (1995) 1-131.}
\nref\EY{B. Eynard, {\it Random Matrices}, Saclay Lecture Notes (2000),
available at {\sl http://www-spht.cea.fr/lectures\_notes.shtml} }
\nref\SCHth{G. Schaeffer, {\it Conjugaison d'arbres
et cartes combinatoires al\'eatoires} PhD Thesis, Universit\'e 
Bordeaux I (1998).}
\nref\CS{P. Chassaing and G. Schaeffer, {\it Random Planar Lattices and 
Integrated SuperBrownian Excursion}, preprint (2002), to appear in 
Probability Theory and Related Fields, math.CO/0205226.}
\nref\GEOD{J. Bouttier, P. Di Francesco and E. Guitter, {\it Geodesic
distance in planar graphs}, to appear in Nucl. Phys. B, cond-mat/0303272.}
\nref\GKY{C. Godr\`eche, I. Kostov and I. Yekutieli, {\it Topological 
Correlations in Cellular Structures and Planar Graph Theory}, Phys.Rev.Lett.
{\bf 69} (1992) 2674-2677; see also D. Boulatov, V. Kazakov, I. Kostov
and A. Migdal, {\it Analytical and Numerical Study of a Model of Dynamically 
Triangulated Random Surfaces}, Nucl. Phys. {\bf B275} (1986) 641-686.}
\nref\DEG{P. Di Francesco, B. Eynard and E. Guitter,
{\it Coloring Random Triangulations},
Nucl. Phys. {\bf B516 [FS]} (1998) 543-587.}
\nref\LALLER{J. Bouttier, P. Di Francesco and E. Guitter, {\it Random
trees between two wall: Exact partition function} Saclay preprint T/03-086 
(2003), cond-mat/0306602.} 
\nref\SCH{G. Schaeffer, {\it Bijective census and random 
generation of Eulerian planar maps}, Electronic
Journal of Combinatorics, vol. {\bf 4} (1997) R20; see also
G. Schaeffer, {\it Conjugaison d'arbres
et cartes combinatoires al\'eatoires} PhD Thesis, Universit\'e 
Bordeaux I (1998).}
\nref\BMS{M. Bousquet-M\'elou and G. Schaeffer,
{\it Enumeration of planar constellations}, Adv. in Applied Math.,
{\bf 24} (2000) 337-368; see also D. Poulalhon and G. Schaeffer,
{\it A note on bipartite Eulerian planar maps}, preprint (2002),
{\sl http://www.loria.fr/$\sim$schaeffe/} and
J. Bouttier, P. Di Francesco and E. Guitter,
{\it Counting colored Random Triangulations},
Nucl.Phys. {\bf B641} (2002) 519-532.}
\nref\CENSUS{J. Bouttier, P. Di Francesco and E. Guitter, {\it Census of planar
maps: from the one-matrix model solution to a combinatorial proof},
Nucl. Phys. {\bf B645}[PM] (2002) 477-499.}

\newsec{Introduction}

The study of random planar graphs is a subject of interest in
several areas of physics, ranging from two-dimensional quantum
gravity, where planar graphs provide a natural discretization
of space-time, to soft condensed matter where they can be
used for instance to model fluid membranes. By random planar
graphs, we mean here graphs embedded in the plane (or
in the two-dimensional sphere), also referred to as fatgraphs
in the physics literature or maps in mathematics.

The statistical properties of random planar graphs have been
studied by various techniques over the years. Beside the original 
combinatorial method due to Tutte \TUT, and the quite technical matrix
integral formulation [\xref\BIPZ-\xref\EY], the most recent approach 
consists in a bijective enumeration \SCHth, based on a one-to-one 
correspondence between
planar graphs and suitably decorated trees, whose statistics
is easily characterized by recursive relations.
Apart from its simplicity, one of the advantages of the bijective approach 
is that it naturally keeps track of the distances between vertices 
on the graphs. This property was used in particular in Ref.\CS\ 
to derive the statistical distribution for the so-called radius of
quadrangulations, and in Ref.\GEOD\ to obtain the distribution of 
distances between two vertices in graphs made of even-sided polygons.
It also gives access to a more refined study of the {\it local environment} 
of a vertex in random graphs, through the statistics of its number of
nearest, next-nearest, ... neighbors. 

The purpose of this paper
is precisely to use the bijective approach to derive the probability 
for a given vertex to have a number, say $n_i$, of neighboring 
vertices and/or a number, say $m_i$, of neighboring edges at a finite 
distance $i$. The case of nearest neighbors was already addressed
in Ref.\GKY\ for the particular case of planar triangulations 
with no multiple edges and no loops. 
In this particular case, a vertex with $k$ neighbors may be replaced by 
a face with $k$ sides, which makes the problem accessible to standard matrix
integral techniques. This approach however breaks down for graphs with 
multiple edges or loops and is limited to the immediate neighbors. 

In this paper, we consider the case of both quadrangulations and
Eulerian (i.e. face bi-colored) triangulations. In the former case,
our analysis relies on a bijection described in Ref.\CS\ between 
rooted planar quadrangulations and so-called well-labeled trees, 
i.e. rooted planar trees with vertices labeled by non-negative integers, 
with a root vertex labeled $1$ and with the constraint that labels on 
adjacent vertices on the tree differ by at most $1$.
In the latter case, a similar bijection will be derived below between
rooted planar Eulerian triangulations and a more restricted class
of well-labeled trees where the labels on adjacent vertices are
required to differ by {\it exactly} $1$.

The paper is organized as follows. 
In Section 2, we present the correspondences between quadrangulations
or Eulerian triangulations on one hand, and well-labeled trees on the
other hand. In Section 3, we first recall known enumeration results 
for well-labeled trees (Sect.3.1), and then show how to extract the statistics of vertices
and/or edges at distance $i$ from a given vertex by computing generating
functions for suitably weighted well-labeled trees (Sect.3.2). When restricting to 
the case of the local environment (i.e. finite distances), we 
extend our study to the limiting case of infinitely large graphs (Sect.3.3).
The following sections (4,5,6) are devoted to a detailed study of
nearest neighbor statistics. In Section 4, we obtain algebraic equations
for the associated generating function, which we use in Section 5
to calculate various probabilities and moments. Section 6 presents 
a heuristic interpretation of our results in the form of a phase diagram for
a statistical model of embedded trees.
We complete our study by presenting in Section 7 a shortcut to treat the 
general case of more distant neighbors. We gather a few concluding remarks in
Section 8.

\newsec{Geodesic distance in planar graphs: From graphs to well-labeled trees}

In this section, we first recall the construction of Ref.\CS\ establishing
a bijection between on one hand rooted planar {\it quadrangulations} and 
on the other hand well-labeled trees, as defined above. This construction
uses crucially the notion of geodesic distance between vertices in the 
quadrangulation and allows for an easy characterization of the number
of neighbors of a given vertex.
We then show how to adapt this construction to establish a bijection
now between rooted planar {\it Eulerian triangulations} and a slightly
different class of well-labeled trees.

\subsec{Quadrangulations}

\fig{A sample rooted planar quadrangulation. An unambiguous representation
in the plane is obtained by having the marked edge adjacent to the external 
face and oriented clockwise around the graph. Each vertex is labeled
by its geodesic distance from the origin of the marked edge (labeled $0$). 
}{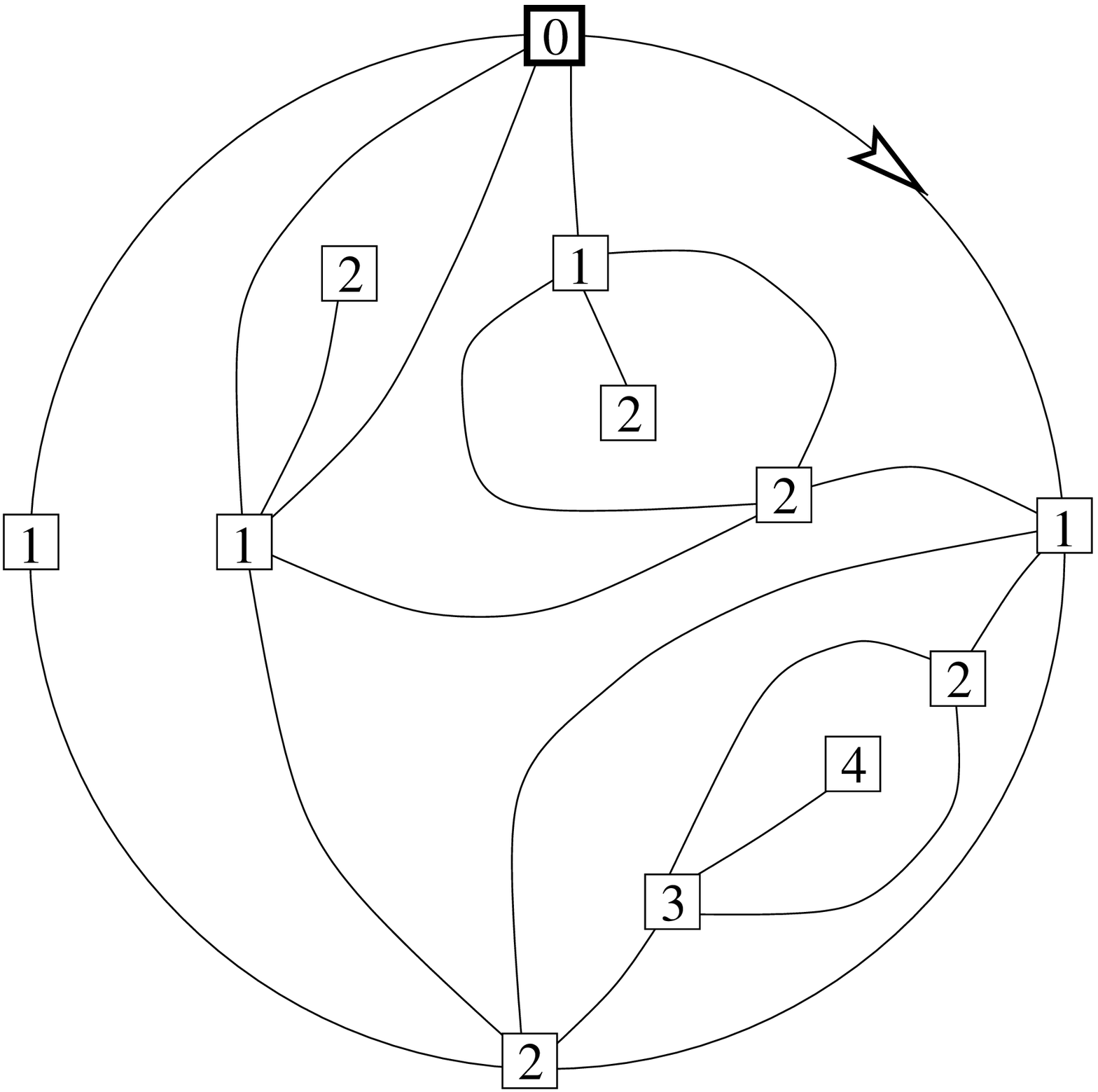}{6.cm}
\figlabel\quadran
Let us start with a rooted planar quadrangulation, i.e a planar map
with all its faces of valence four (i.e. having four sides) and with a marked
edge bordering the infinite face and oriented clockwise around
the quadrangulation. 
The {\it origin} of the quadrangulation is the vertex from which 
the marked edge originates. For each vertex, we define its geodesic distance 
from the origin as the minimal number of edges of the paths linking the origin 
to this vertex. We may then label each vertex by its geodesic distance from the
origin. With this labeling, the origin is the only vertex labeled by $0$, while
all other vertices have positive labels. In particular, the nearest 
neighbors of the origin are the vertices labeled by $1$ (see Fig.\quadran). 
\fig{The two possible configurations of labels around a face in a 
quadrangulation: (i) a simple face and (ii) a confluent face. To each face
is associated an edge (thick solid line) as indicated below. The collection
of these edges will eventually produce the well-labeled tree associated
to the original quadrangulation.
}{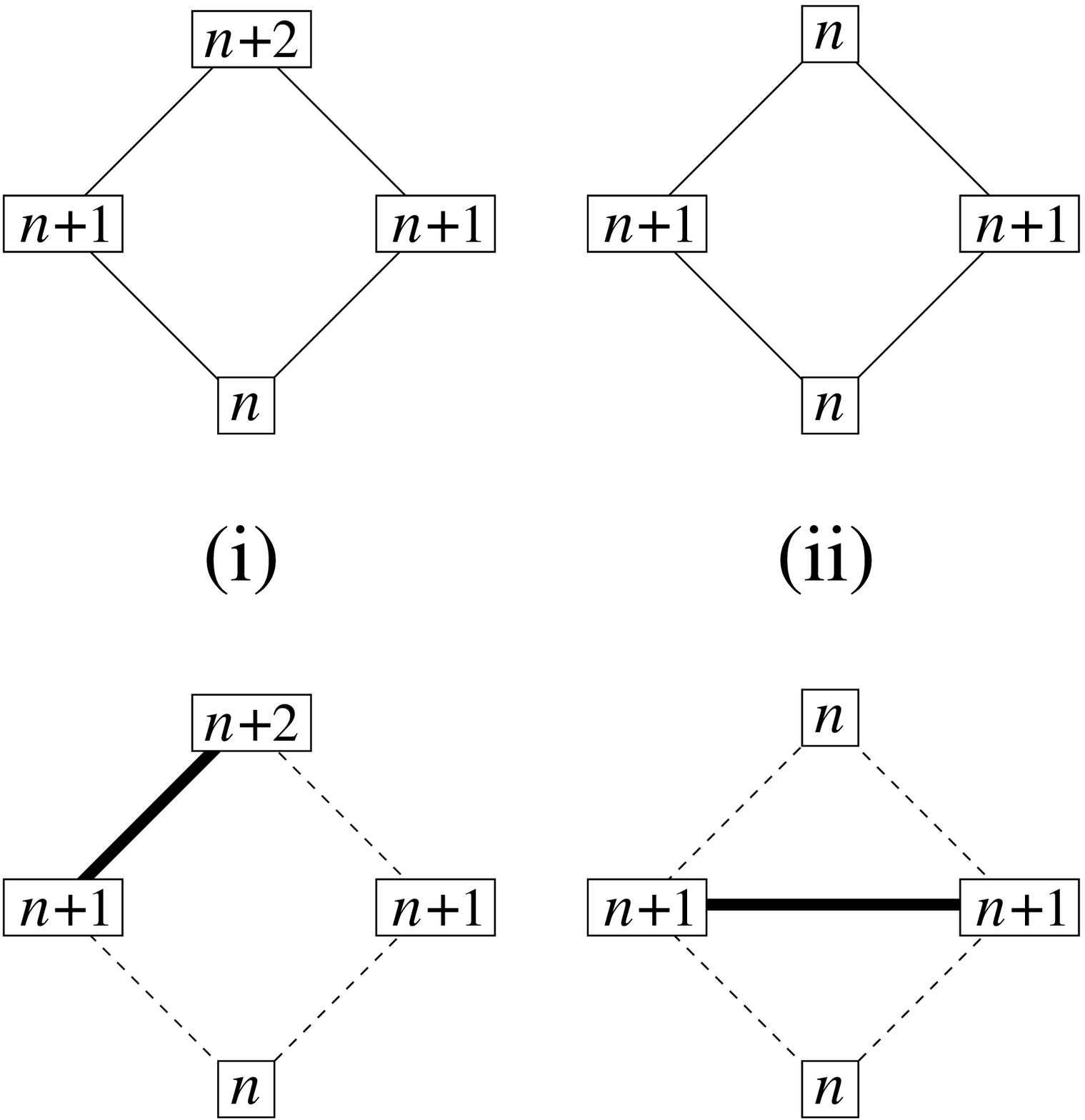}{7.cm}
\figlabel\rulequa
Each face of the quadrangulation may be classified, upon reading
the four labels of the vertices around it in clockwise direction, in 
one of the two following categories (see Fig.\rulequa) :
\item{(i)} ``simple'' faces with a sequence of labels of the form  
$n\to n+1 \to n+2 \to n+1$
\item{(ii)} ``confluent'' faces with a sequence of labels of the form 
$n\to n+1\to n\to n+1$
\par
Indeed, neighboring vertices have labels which differ by at most one
by construction and the bipartite nature of the quadrangulation 
implies that the parity of the labels changes from any vertex to 
its neighbor.
We now build a well-labeled tree with rules resulting from the above
classification (see Fig.\rulequa):
\item{(i)} For each simple face, we retain the edge $n+1\to n+2$ 
(in clockwise direction
\foot{Here and throughout the paper,
when we work in the planar representation, the rule must be reversed
for the external face, i.e. the labels are to be read in {\it counterclockwise}
direction around the graph.}) 
\item{(ii)} For each confluent face, we draw a new edge linking the
two vertices with labels $n+1$
\par
\fig{The impossibility of loops in the collections of edges as built
from the rules of Fig.\rulequa. Indeed, such a loop would have
a vertex with minimal label $n$ whose environment would be either (a) or (b).
In both cases, a vertex of label $n-1$ is present on both sides
of the loop, which contradicts the definition of labels as
geodesic distances from the origin.}{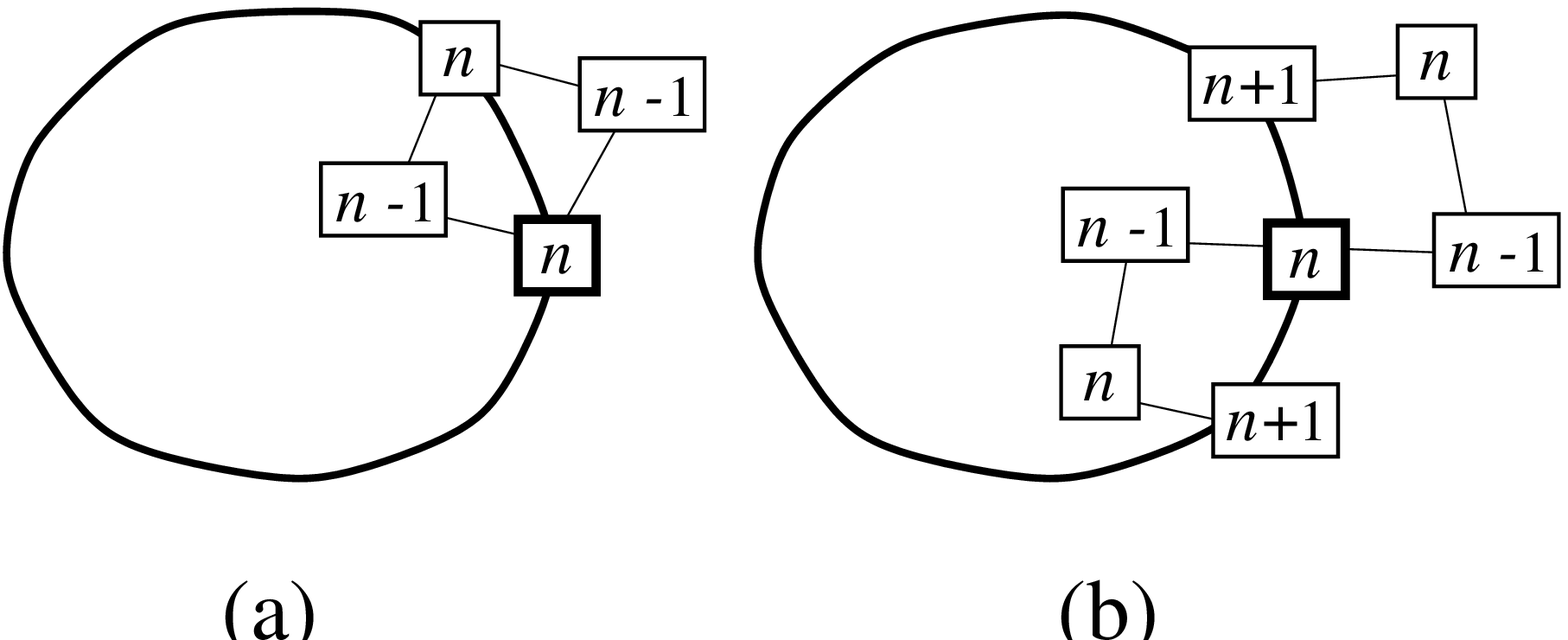}{9.cm}
\figlabel\loopqua
The object resulting from the collection of these edges is a planar graph 
whose vertices are vertices of the original quadrangulation. 
Let us now show that it is indeed a well-labeled tree. 
Let $N$ denote the number of faces
of the original quadrangulation, with therefore $2N$ edges and  
$N+2$ vertices. We note that $N$ is also the number of
edges of the graph resulting from the above construction as 
each face produces an edge and
the resulting edges are all distinct (this is clear for confluent faces
and results from the orientation prescription for simple faces).
It is now sufficient to prove that the graph contains no loop.
Suppose that such a loop exists, we pick a vertex with a minimal
label, say $n$, along this loop. By examining all possible configurations 
(see Fig.\loopqua), we see that this vertex is adjacent to at least
one vertex labeled $n-1$ on each side of the loop. This is a contradiction
as the labels represent the geodesic distance: among the geodesic paths  
linking the origin to these vertices, one must cross the loop at
a vertex with label less than $n-1$, in conflict with the property 
that all labels on the loop are larger than $n$.
The graph is therefore made of $C\geq 1$ connected planar trees, hence 
has $N+C$ vertices. On the other hand, the number of vertices
is at most $N+1$ as the origin cannot belong to a retained edge.
We deduce that $C=1$, hence the graph is a connected tree containing 
all the vertices of the quadrangulation but the origin. This tree
is well-labeled as all labels are positive and labels
on adjacent vertices differ by at most one by construction.
\fig{The construction of the well-labeled tree associated with the
rooted quadrangulation of Fig.\quadran: the edges
selected (a) according to the rules of Fig.\rulequa\ produce 
the well-labeled tree (b). The tree is rooted at the vertex
(labeled $1$) where the original marked edge ends. We have shaded the confluent
faces in (a).}{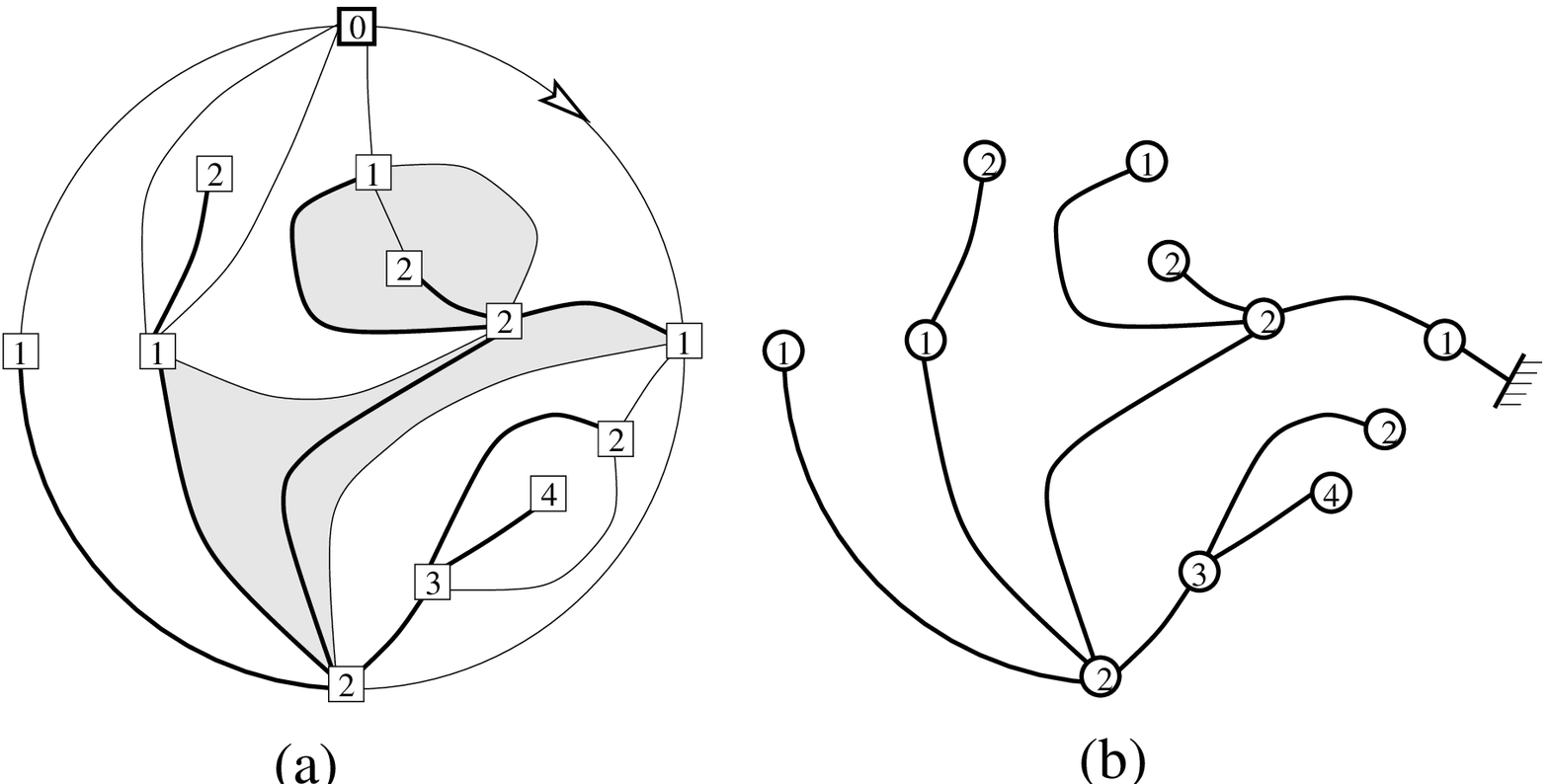}{12.cm}
\figlabel\quadtree
\noindent The tree is naturally rooted according to the position of the 
marked edge in the original rooted quadrangulation (see Fig.\quadtree)
\fig{The inverse construction of Fig.\quadtree\ 
from the well-labeled tree (a) to the quadrangulation of Fig.\quadran.
We first restore the origin and connect it to each corner at all vertices labeled
$1$ (b). Within each of the newly created faces, further connections are
made according to the rules explained in the text (c). The final step
is to remove all edges connecting vertices with equal labels (d).}{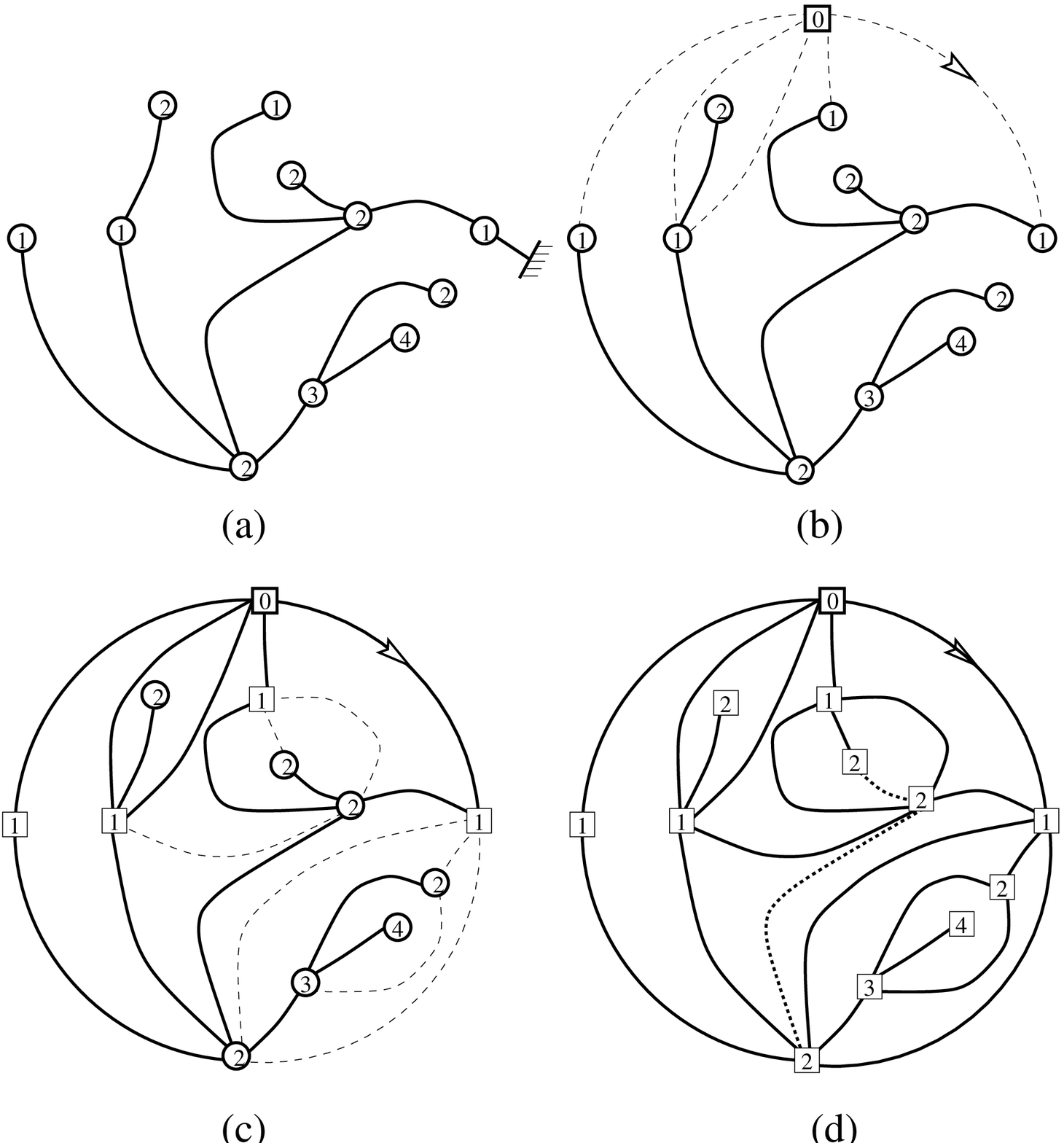}{12.cm}
\figlabel\treequad
So far we have associated to each rooted planar quadrangulation
a rooted planar well-labeled tree. 
Let us now describe the inverse procedure.
Starting from a well-labeled tree, we restore an origin by adding 
an extra vertex labeled $0$. Non-intersecting edges are added between 
this vertex and each vertex labeled $1$ with one connection per {\it corner} 
at this vertex (see Fig.\treequad). The position of the root of the tree
naturally translates into a marking of one of these newly created edges.
The resulting map is made of a number of faces in which we apply
the following construction independently. Reading the labels of the corners 
(as inherited from the label of the adjacent vertex) in clockwise direction 
around the face, 
we connect by a new edge each corner with label $n\geq 2$ 
to the first encountered corner with label $n-1$, unless this latter
corner is the immediate successor of the former and in such a way
that the added edges do not cross each other. Finally, a planar rooted 
quadrangulation is obtained by erasing all the edges linking vertices
with identical labels.
We rely on Ref.\CS\ for a formal proof that the above constructions
are inverse of one-another. 

This bijection between rooted planar quadrangulations and well-labeled
trees keeps track of the local environment of the origin. For
instance, the vertices labeled $1$ in the tree are
the nearest neighbors of the origin, those labeled $2$ the next-nearest
neighbors and so on. This allows in particular to count the number
of nearest neighbors of the origin as the number of vertices labeled
$1$ in the tree. Note that this number {\it does not} account for
the multiplicity of the neighbors, which are all counted once even if
they are connected by several edges to the origin. 
The number of nearest neighbors counted with
their multiplicity is nothing but that of edges connected to the origin. 
In the language of well-labeled trees, this is the total number of corners 
labeled $1$, or equivalently the sum of the valences of all vertices labeled 
$1$.
 
\subsec{Eulerian triangulations}

\fig{A sample rooted planar Eulerian (face-bicolored) triangulation. An unambiguous 
representation in the plane is obtained by having the marked edge adjacent 
to the external face and oriented clockwise around the graph. This orientation
induces an orientation of all the other edges as shown. Each vertex is labeled
by its geodesic distance from the origin of the marked edge (labeled $0$).
This distance is now defined with paths respecting the orientation.
}{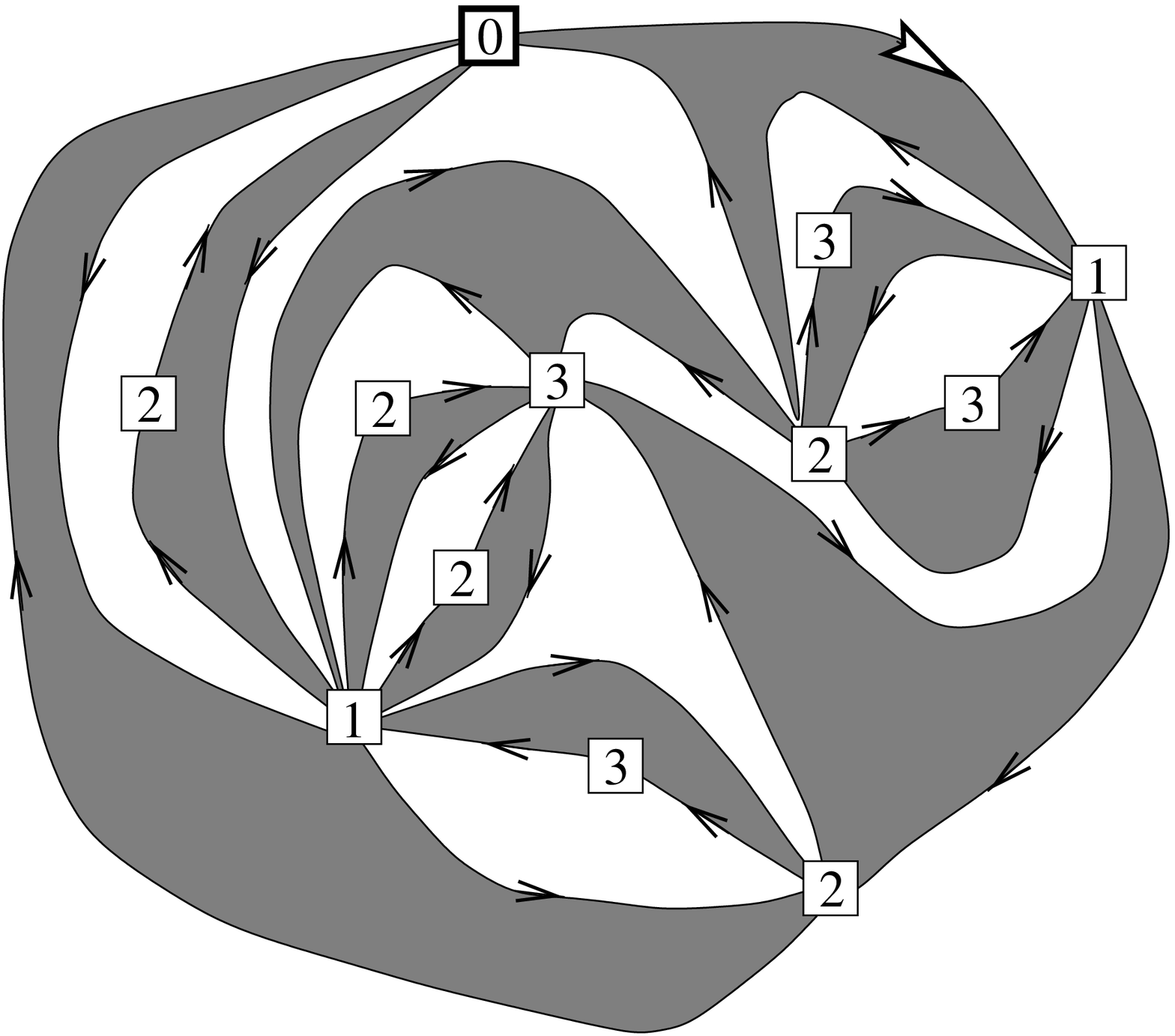}{7.cm}
\figlabel\trian
We now intend to adapt the above constructions to the case of planar 
{\it Eulerian} triangulations, i.e. planar maps with all
faces of valence three (hereafter referred to as triangles) and with an even
number of triangles around each vertex (see Fig.\trian). These triangulations
are equivalently characterized by the property that their
faces are bicolorable or that their vertices are tricolorable
\DEG. 
\fig{The configuration of labels around a clockwise oriented (black) face in 
an Eulerian triangulation. To each black face is associated an edge 
(thick solid line) as indicated on the right. The collection
of these edges will eventually produce the well-labeled tree associated
to the original Eulerian triangulation.
}{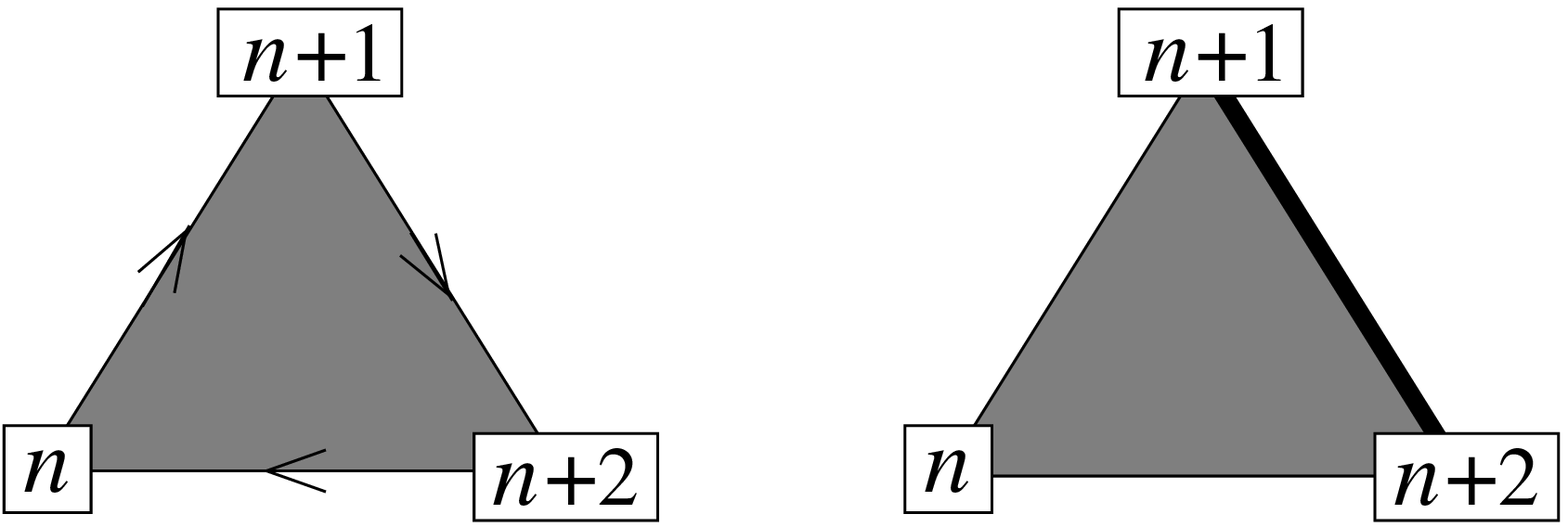}{9.cm}
\figlabel\ruletri
We start from a rooted such 
triangulation, with a marked edge bordering the infinite face and
oriented clockwise around the triangulation. This fixes a natural 
orientation for every edge by requiring that orientations alternate
around each vertex. This in turn gives a natural bicoloration of the
faces in say black and white according to the clockwise or counterclockwise
orientation of the adjacent edges. Taking as before for origin the vertex 
from which the marked edge originates, we define the geodesic distance
of any vertex from the origin as the minimal number of edges of
the paths linking the origin to that vertex and {\it respecting the
orientations}. This defines a labeling of the vertices, such that the
sequence of labels around all clockwise oriented (black) faces is
of the form $n\to n+1\to n+2$ (see Fig.\ruletri). This follows
from the fact that neighboring vertices have labels which differ by
at most two while the tricolorability of the vertices fixes
the residue modulo $3$ of all labels.
We now build a well-labeled tree by a similar rule as for quadrangulations
by retaining for each clockwise oriented (black) face with labels
say $n\to n+1\to n+2$ the edge linking the vertices labeled $n+1$ and $n+2$
(see Fig.\ruletri).
\fig{The impossibility of loops in the collections of edges as built
from the rules of Fig.\ruletri. Indeed, such a loop would have
a vertex with minimal label $n$ which would be adjacent to
a vertex labeled $n-1$ on both sides
of the loop, which contradicts the definition of labels as
geodesic distances from the origin.}{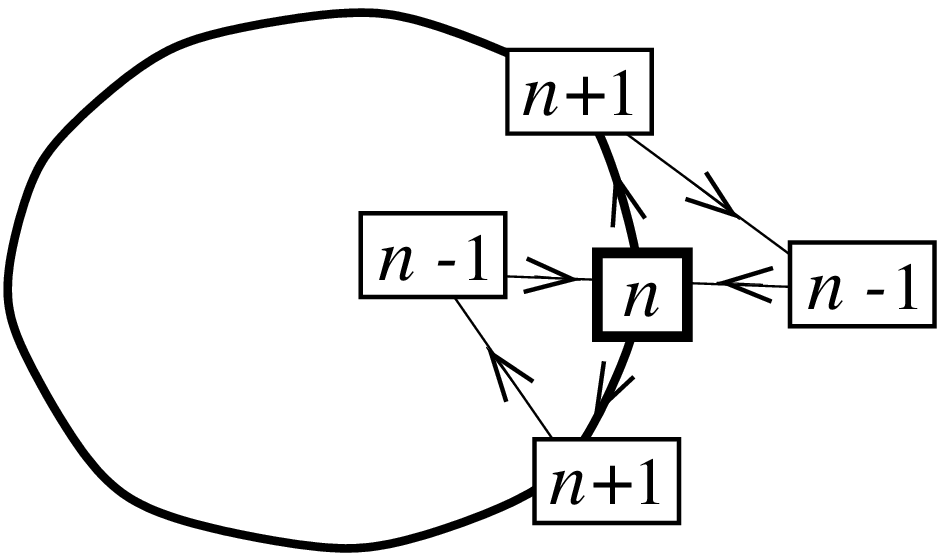}{5.cm}
\figlabel\looptri
The resulting object is a well-labeled tree with the stronger constraint
that adjacent vertices have labels differing by {\it exactly} $1$.
To see that, we follow the same line of reasoning as above. 
Let $N$ denote the number of black faces of the triangulations
with therefore $2N$ faces, $3N$ edges and $N+2$ vertices. The
graph resulting from the above construction has $N$ edges and contains 
no loop (see Fig.\looptri).
It is therefore made of $C$ connected trees and $C=1$ since
the total number of vertices $N+C$ is less than $N+1$ as the
origin does not belong to any selected edge.
\fig{The construction of the well-labeled tree associated with the
rooted Eulerian triangulation of Fig.\trian: the edges
selected (a) according to the rules of Fig.\ruletri\ build
the well-labeled tree (b)}{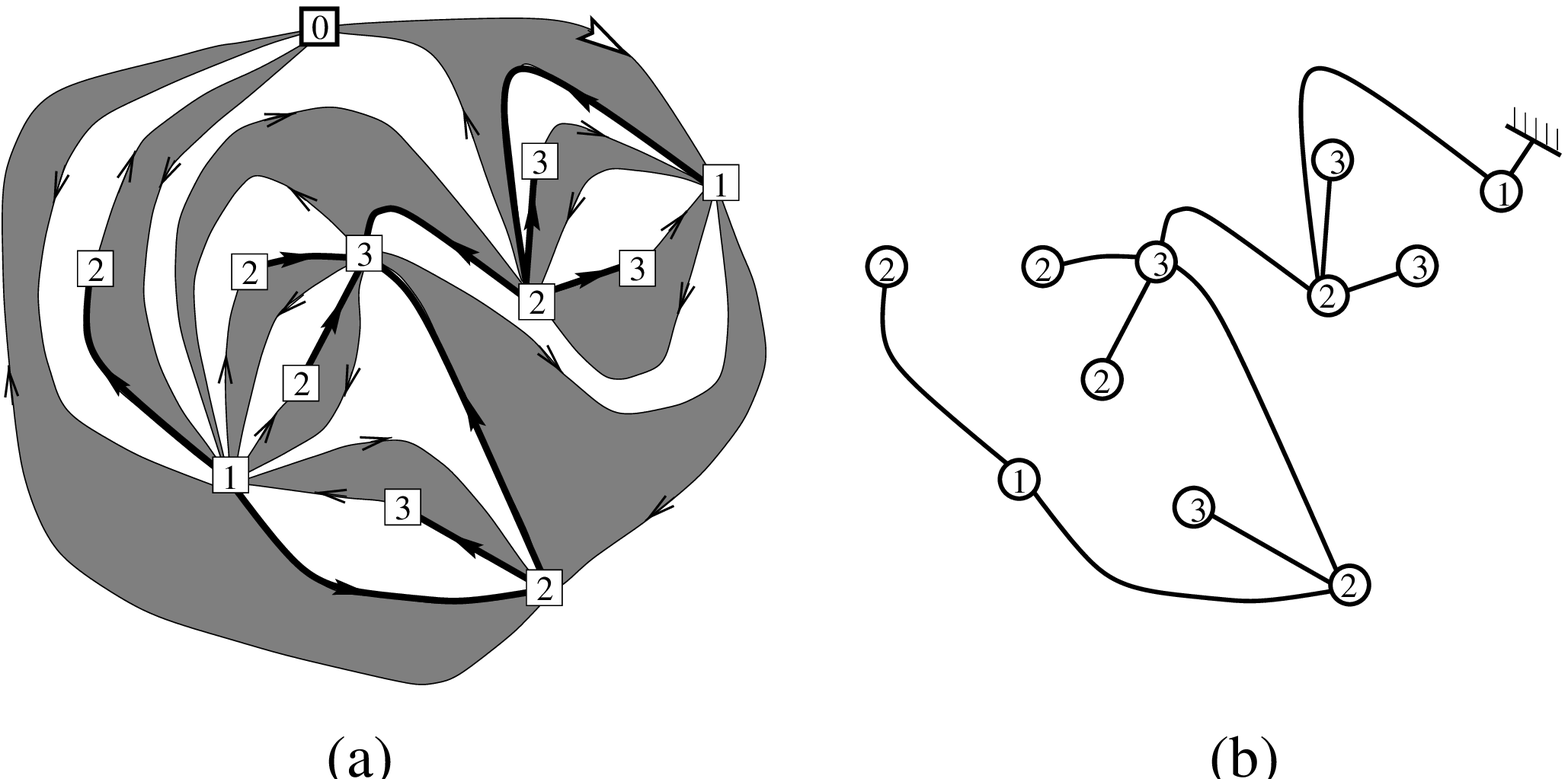}{12.cm}
\figlabel\tritree
\noindent The tree is naturally rooted according to the position of 
the marked edge
in the original rooted Eulerian triangulation (see Fig.\tritree)
\fig{The inverse construction of Fig.\tritree. Starting from
the well-labeled tree, we first restore the origin and link it to the
right of each $1\to 2$ edge (a). The links are then inflated into 
$0\to 1\to 2$ black
triangles (b). We proceed within each of the newly created white faces 
by linking the unique vertex labeled $1$ within the face to the
right of the $2\to 3$ edges (c). These links are inflated into $1\to 2\to 3$
black triangles. The procedure is repeated until an Eulerian
triangulation is obtained, which is already the case here (d).
}{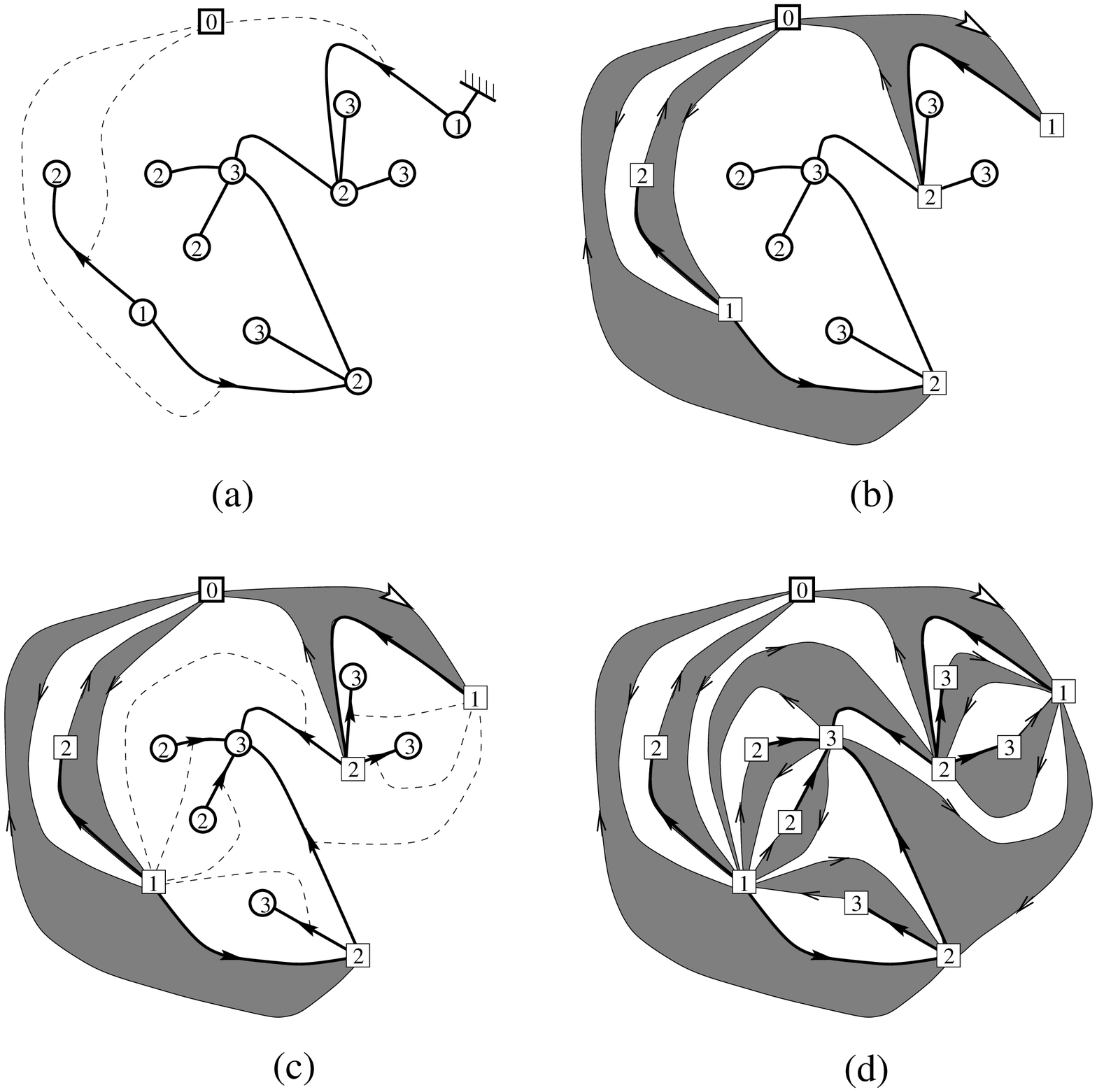}{12.cm}
\figlabel\treetri
Let us now describe the inverse construction, which consists
in iteratively building the black clockwise-oriented triangles.
Starting from a rooted well-labeled tree with adjacent labels
differing by exactly one, we construct the associated Eulerian
triangulation step by step as follows.
The step $0$ consists in adding an origin and linking it to
the right of each edge with labels $1$ and $2$, oriented
from $1$ to $2$. Each such link is inflated so as to create
a clockwise oriented (black) triangle $0\to 1\to 2$. Beside these
triangles, the original external face is split into a number of 
white faces (see Fig.\treetri). It is easily seen that the sequence of
labels in clockwise orientation around any of these white faces
is of the form $0\to 2\to \cdots\to 2\to 1$ where all labels
in-between the first and the last $2$-labels are greater or equal
to $2$ and all increments but the first ($0\to 2$) are $\pm 1$.
In particular, if the sequence is reduced to $0\to 2\to 1$, a
white triangle has been created. We then proceed to the step
$1$ which takes place independently inside each white face not
reduced to a triangle. For a given such face $F$,  
each $2\to 3$ edge having its right side adjacent to $F$   
is linked from this side to the unique vertex labeled $1$ around $F$.
As before, we inflate this link, thus creating a black triangle 
$1\to 2\to 3$. As $F$ is not reduced to a triangle, it is easy to
see that at least one black triangle is created. Beside these
additional black faces, the face $F$ is split into smaller
white faces. Among these faces, that containing the origin
is a triangle $0\to 2\to 1$ and the others of the form
$1\to 3\to \cdots\to 3\to 2$ such that all labels
in-between the first and the last $3$-labels are greater or equal
to $3$, and all increments but the first ($1\to 3$) are $\pm 1$.
These faces, if not reduced to triangles, have to be split again by 
repeating the above procedure.
At step $i$, all faces are either black or white triangles 
or larger white faces of the form $i\to i+2\to \cdots\to i+2\to i+1$
where all labels
in-between the first and the last $i+2$-labels are greater or equal
to $i+2$ and all increments but the first ($i\to i+2$) are $\pm 1$.
The procedure ends when all labels are exhausted and leaves us with
an Eulerian triangulation rooted at the $0\to 1$ edge linking the
origin to the root of the original tree.

Let us now finally show that the two constructions above,
respectively from an Eulerian triangulation to a well-labeled tree 
and conversely are inverse of one-another. It is straightforward
that, starting from a well-labeled tree and building the associated
triangulation, the latter is re-decomposed into the same labeled
tree as the construction of black triangles always consists
in adding edges of the form $n\to n+1$ and $n+2\to n$ to 
tree-edges of the form $n+1\to n+2$.  To complete the
proof of the bijection it is now sufficient to show that the number of 
objects is the same within the two classes, namely on one hand rooted
planar Eulerian triangulations with $N$ black faces, and on the other hand
rooted well-labeled trees with $N$ edges. The former number
was first obtained in Ref.\TUT\ with the result 
$3\, 2^{N-1} (2N)!/(N!(N+2)!)$. The latter number will be 
derived below and is identical, which completes the proof.

As for the case of quadrangulations, the bijection between Eulerian
triangulations and well-labeled trees allows to keep track of 
the numbers of nearest, next nearest, etc ... neighbors of the
origin in Eulerian triangulations. Note that the notion of
neighbors now uses the edge orientations. As before, we have access
to the number of neighbors both with and without their multiplicity.

\newsec{Planar graphs viewed from a vertex: statistics of neighboring edges
and vertices}

In this section, we use the above bijections to study the general statistics
of edges and vertices at fixed geodesic distances from a given vertex (origin)
in arbitrary quadrangulations and Eulerian triangulations. 
This is done via the enumeration of properly weighted well-labeled
trees. We first recall how to perform this enumeration by use of recursion
relations in the absence of weights, and then proceed to the weighted
case. Finally, in the case of neighbors within a finite range of distances,
we show how to derive algebraic equations for the various generating 
functions at hand. 

\subsec{Enumeration of well-labeled trees}

The above bijections reduce the problem of enumeration of
quadrangulations or Eulerian triangulation to that, much simpler,
of well-labeled trees. 
This is done via 
the computation of their generating functions with a
weight $g$ per edge. More precisely, we introduce 
the generating functions $W_j$, $j=1,2,\cdots$, for rooted trees 
with positive labels and with a root labeled $j$.
In view of the above bijections, the function $W_1$ is indeed identified with
the generating function for rooted quadrangulations 
(resp. Eulerian triangulations) with a weight $g$ per face 
(resp. per black face). We are therefore only interested in $W_1$, but
in practice, the scheme we propose involves the
simultaneous computation of all $W_j$'s. From the constraint that adjacent 
vertices have labels differing by at most or exactly one, we
deduce the following master equations:
\eqn\master{\eqalign{W_j&={1\over 1-g(W_{j+1}+W_j+W_{j-1})}\qquad 
{\rm (Quadrangulations)}\cr W_j&={1\over 1-g(W_{j+1}+W_{j-1})}\qquad 
{\rm (Eulerian\ Triangulations)}\cr}}
obtained by inspection of the descendents of the root vertex (see Ref.\LALLER). 
These recursion relations must be supplemented by the boundary condition
$W_0=0$ and restricted to values of $j=1,2,3,\cdots$. The solution is
then unique as one readily sees by expanding eqs.\master\ order by
order in $g$ and noting that all $W$'s are 
series of $g$ with only non-negative powers by definition.

The explicit solutions of eqs.\master\ were worked out in Ref.\GEOD,
with the results
\eqn\results{\eqalign{W_j&=W {u_j u_{j+3} \over u_{j+1} u_{j+2}}\qquad
{\rm (Q.)}\cr W_j&=W {u_j u_{j+4} \over u_{j+1} u_{j+3}}\qquad
{\rm (E.T.)}\cr}}
with 
\eqn\formW{\eqalign{W&= {1-\sqrt{1-12g}\over 6g}\qquad
{\rm (Q.)}\cr W&={1-\sqrt{1-8g}\over 4g} \qquad
{\rm (E.T.)}\cr}}
and where
\eqn\formu{u_j=1-x^j\quad {\rm with} \quad \left\{\matrix{x= 
{1+2 \theta^2 -\theta \sqrt{3(2+\theta^2)}\over 1-\theta^2} & \theta=(1-12 g)^{1\over 4} &
{\rm (Q.)}\cr x= {1-\theta\over 1+\theta} & \theta=(1-8 g)^{1\over 4} &
{\rm (E.T.)}\cr}\right.}
In particular, eqs.\results\ yield the following formulas for $W_1$:
\eqn\forWone{\eqalign{
W_1&= {4(1+2\theta^2)\over 3(1+\theta^2)^2} \qquad {\rm (Q.)} \cr
W_1&={5+10 \theta^2+\theta^4\over 4(1+\theta^2)^2} \qquad {\rm (E.T.)} \cr}}
with $\theta$ as in eq.\formu.
For Eulerian triangulations, we easily extract from the above formula
the number of well-labeled trees with $N$ edges 
\eqn\extWone{ W_1\vert_{g^N}= 3\, 2^{N-1} {(2N)!\over N!(N+2)!} }
where the subscript $\vert_{g^N}$ refers to the coefficient of $g^N$ in
the corresponding series. 
This number coincides with the number of rooted planar Eulerian triangulations
with $2N$ faces, and allows to complete the proof of the bijection of Sect.2,
as announced.

The function $W$ of eq.\formW\ is the limit of $W_j$ for large $j$. By a 
global shift of labels, the function $W_j$ may be understood as counting 
trees with a root labeled $1$ and with all labels $\geq 2-j$. The function
$W$ therefore simply counts rooted trees with a root labeled $1$ and with  
possibly negative integer labels. Interestingly enough, the function
$W$ may also be interpreted in the language of graphs.
In the case of quadrangulations, it is easily identified as 
the generating function with a marked vertex and a marked edge, while
$W_1$ is the generating function with only a marked oriented edge (here
we make use of the spherical representation of planar graphs for simplicity).
If $Q_N$ denotes the total number of quadrangulations on the sphere 
with $N$ faces (counted with their inverse symmetry factor), 
we have $W|_{g^N}=(N+2)(2N)Q_N$ while $W_1|_{g^N}=2(2N)Q_N$, 
hence $W_1|_{g^N}=2/(N+2)W|_{g^N}=2\,3^{N}c_N/(N+2)$ 
where $c_N={2N\choose N}/(N+1)$ is the $N$-th Catalan number,
a result already obtained in Ref.\SCH\ via tree conjugacy, and consistent
with the solution \forWone, as checked in Ref.\GEOD.

Similarly, in the case of Eulerian triangulations, the function $W$
is identified with the generating function with a marked (origin) vertex and 
a marked oriented edge of type $j\to j+1$ with respect to this origin 
and with a global orientation induced by that of the marked edge, while
$W_1$ is the generating function with a marked oriented edge (here again,
we make use of the spherical representation of planar graphs for simplicity).
If $E_N$ denotes the total number of oriented Eulerian triangulations 
on the sphere with $2N$ faces (counted with their inverse symmetry factor), 
we have $W|_{g^N}=(N+2)(2/3)(3N)E_N$, as for a fixed origin, exactly $2/3$ 
of the edges will be of type $j\to j+1$, while $W_1|_{g^N}=(3N)E_N$, 
hence $W_1|_{g^N}=3/(2(N+2))W|_{g^N}=3\, 2^{N-1}c_N/(N+2)$,
a result already obtained in Ref.\DEG\ via matrix models and in Ref.\BMS\ 
via tree conjugacy, and also consistent with the solution \extWone.

\subsec{Statistics of edges and vertices at fixed distances from an origin}

So far the counting functions have not kept track of
the numbers of vertices and edges at a given distance from the origin
in the graphs. This can be repaired
by considering the same generating functions $W_j$ 
as before but with extra weights,
namely $\rho_i$ per vertex labeled $i$ in the tree, and
$\alpha_i$ per edge adjacent to a vertex labeled $i$ 
(an edge linking a vertex labeled $i$ to a vertex labeled $k$
thus receives the weight $\alpha_i\alpha_k$). From now on, 
we shall use the shorthand notation 
$W_j\equiv W_j(\{\alpha_i;\rho_i\})$ for the generating function
incorporating all the weights $\alpha_i$ and $\rho_i$, while the $W_j$'s
of previous section will be written as $W_j(\{1;1\})$.
Taking into account these new weights, the master equations become 
\eqn\newmaster{\eqalign{W_j&={\rho_j\over 1-g\alpha_j(\alpha_{j+1}W_{j+1}+
\alpha_j W_j+\alpha_{j-1}W_{j-1})}\qquad 
{\rm (Q.)}\cr W_j&={\rho_j\over 1-g\alpha_j(\alpha_{j+1}W_{j+1}+
\alpha_{j-1}W_{j-1})}\qquad 
{\rm (E.T.)}\cr}}
which, together with $W_0=0$, determine again the
$W$'s completely as powers series in $g$. 

Let us now show how to extract from these generating functions 
the statistics of neighbors in quadrangulations of Eulerian triangulations.
The neighbor statistics is entirely characterized by the following average:
\eqn\statavN{\eqalign{\Gamma_N & \equiv\langle \prod_{j\geq 1} \alpha_j^{m_j}
\rho_j^{n_j}\rangle_N = {\Delta_N(\{\alpha_j;\rho_j\})\over \Delta_N(\{1;1\})}\cr
\Delta_N(\{\alpha_j;\rho_j\}) & =\sum_{({\cal G}, v\in {\cal G})} 
\alpha_j^{m_j({\cal G},v)}
\rho_j^{n_j({\cal G},v)}/|Aut({\cal G},v)|\cr}}
where the statistical sums extend over all pairs $({\cal G},v)$ made of 
a quadrangulation ${\cal G}$ (resp. an Eulerian triangulations ${\cal G}$) 
with $N$ faces (resp. $N$ black faces) on the sphere and a vertex 
$v$ of ${\cal G}$.
Here $n_j({\cal G},v)$ is the number of vertices at geodesic distance
$j$ from the vertex $v$ while  $m_j({\cal G},v)$ denotes the number
of edges $(j-1,j)$ linking  a vertex at distance $j-1$ from $v$
to a vertex at distance $j$ from $v$. 
For instance, $n_1({\cal G},v)$ (resp. $m_1({\cal G},v)$) denotes 
the number of nearest neighboring vertices (resp. edges) of $v$ in $\cal G$.
The factor $|Aut({\cal G},v)|$
is the usual symmetry factor of the marked graph on the sphere.

The statistical sum $\Delta_N$ is related to $W_1$ via
\eqn\Wdel{\alpha_1{d\over d\alpha_1} \Delta_N(\{\alpha_j;\rho_j\})=
W_1(\{\alpha_j;\rho_j\})|_{g^N}}
The derivation with respect to $\alpha_1$
amounts to counting the graphs ${\cal G}$ (already marked at $v$) with a multiplicity 
$m_1({\cal G},v)$ equal to the number of edges originating from $v$.
This corresponds precisely to counting graphs with a marked
oriented edge on the sphere, i.e. {\it rooted} planar maps. 
\fig{The one-to-one correspondence between $(j-1,j)$ edges (represented in thick
lines)
in quadrangulations (top) or Eulerian triangulations (bottom) and edges
adjacent to vertices labeled $j$ in the corresponding well-labeled trees
(dashed lines).}{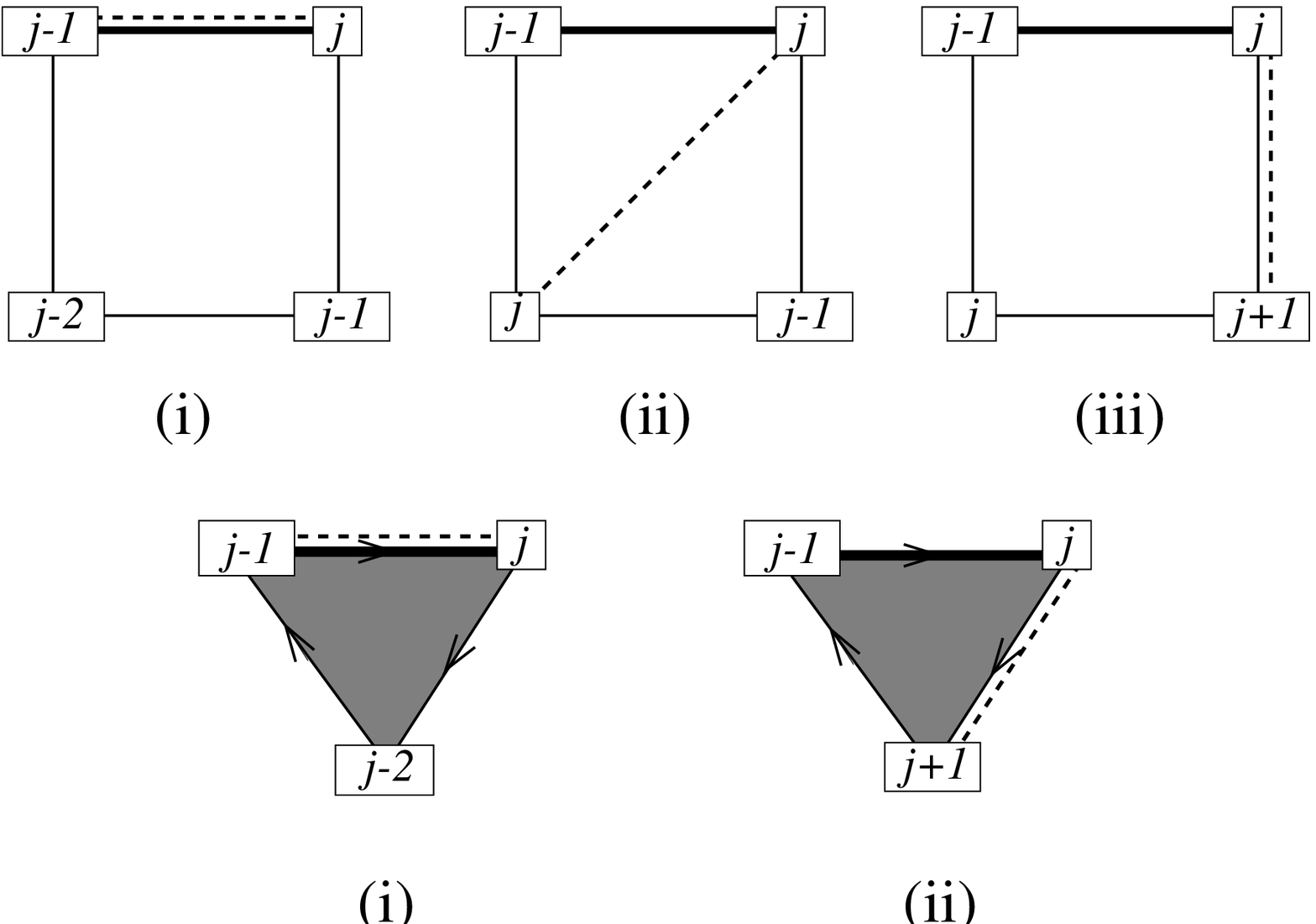}{12.cm}
\figlabel\inspec
The relation \Wdel\ follows from the fact that the total number of vertices at geodesic
distance $j$ from the origin is equal to that of vertices labeled $j$ in the
associated well-labeled trees, while
the total number of edges $(j-1,j)$ in the graph
is equal to the total number of edges adjacent to a 
vertex labeled $j$ in the associated well-labeled tree. 
To see the latter property, note that the face
adjacent to the $(j-1,j)$ edge and on its right contains 
exactly one edge of the tree adjacent to the vertex labeled
$j$ in this edge. This is done by inspecting all possible cases for 
quadrangulations and Eulerian triangulations (see Fig.\inspec). 
Conversely, to each edge adjacent to a vertex labeled $j$ in the well-labeled 
tree, say of the form $(j,i)$, we may associate a unique edge $(j-1,j)$ 
of the graph.
We have indeed the following possibilities for quadrangulations (respectively
associated to the three situations of the first line of Fig.\inspec):
\item{(i)} $i=j-1$: the edge $(i,j)$ is the $(j-1,j)$ edge
\item{(ii)} $i=j$: the edge $(i,j)$ is a diagonal of a square
$(j-1,j,j-1,j)$, and we pick the edge $(j-1,j)$ on its left
\item{(iii)} $i=j+1$: the edge $(j,i)$ has a square on its right
of the form $(j-1,j,j+1,j)$, in which we pick the edge $(j-1,j)$ adjacent
to $(j,i)$
\par
\noindent For Eulerian triangulations we only have two possibilities
(respectively
associated to the two situations of the second line of Fig.\inspec):
\item{(i)} $i=j-1$: the edge $(i,j)$ is the $(j-1,j)$ edge
\item{(ii)} $i=j+1$: the edge $(j,i)$ is adjacent to
a black triangle on its right, of the form $(j-1,j,j+1)$, 
in which we pick the $(j-1,j)$ edge
\par
\noindent This completes the bijection between the edges $(j-1,j)$ of the graph
and those adjacent to the same vertex labeled $j$ in the tree.

In conclusion, inverting eq.\Wdel, the knowledge of $W_1$ as determined by eqs.\newmaster\
gives access to the statistical average $\Gamma_N$ via
\eqn\invstat{ \Gamma_N= {\int_0^{\alpha_1} {d\alpha \over \alpha}
W_1(\{\alpha,\alpha_2,\cdots;\rho_1,\rho_2,\cdots\})\vert_{g^N}\over
\int_0^1{d\alpha \over \alpha}
W_1(\{\alpha,1,\cdots;1,1,\cdots\})\vert_{g^N}}}

\subsec{General solution for finite distances}

{}From now on, we will concentrate on the study of the {\it local} environment
of a given vertex. By this, we mean the statistics of edges
and vertices within a {\it finite} range, say $k$, of geodesic distances
from an origin. To this end, we may limit ourselves to a finite
set of parameters $\{\alpha_1,\cdots,\alpha_k;\rho_1,\cdots \rho_k\}$
while all other $\alpha$'s and $\rho$'s are taken equal to one.
Here we wish to present a very general scheme for solving this
problem in the form of algebraic equations for the generating 
functions.

Introducing the functions
\eqn\ZW{Z_j=\alpha_j W_j}
the equations \newmaster\ split into two sets of equations:
a finite set
\eqn\splitmaster{\eqalign{Z_j&={\alpha_j\rho_j\over 
1-g\alpha_j(Z_{j+1}+ Z_j+Z_{j-1})}\qquad {\rm (Q.)}\cr 
Z_j&={\alpha_j\rho_j\over 
1-g\alpha_j(Z_{j+1}+Z_{j-1})} \qquad {\rm (E.T.)}\cr}}
valid for $1\leq j\leq k$, and an infinite set 
\eqn\mastersplit{\eqalign{Z_j&=
{1\over 1-g(Z_{j+1}+ Z_j+Z_{j-1})}\qquad {\rm (Q.)}\cr 
Z_j&= {1\over 1-g(Z_{j+1}+Z_{j-1})}\qquad {\rm (E.T.)}\cr}}
for all $j\geq k+1$.
Remarkably enough, this latter infinite set of equations may be 
replaced by a {\it single equation}. Indeed, these equations may be solved
order by order in $g$ from the initial data $Z_k$.  
Moreover, it is clear from the interpretation of the $Z_j=W_j$ for
$j\geq k+1$ that $Z_j\to W$ as in \formW\ when $j\to \infty$.
This is equivalently ensured by a single relation between
$Z_{k+1}$ and $Z_k$ thanks to the existence of a {\it discrete integral
of motion} $f(Z_j,Z_{j+1})=f(Z_{j+1},Z_{j+2})$ $j\geq k$, with
\eqn\intmo{\eqalign{
f(x,y)&= x y (1-g(x+y)) -x-y \qquad {\rm (Q.)}\cr
f(x,y)&= x y (1-gx)(1-gy)+g x y -x-y \qquad {\rm (E.T.)}\cr}}
The convergence is ensured by writing 
\eqn\intmagic{f(Z_k,Z_{k+1})=f(W,W)}
This equation is now supplemented by the finite set \splitmaster\
of $k$ equations, and upon elimination we may write a single
algebraic equation for say $W_1=Z_1/\alpha_1$.

As an illustration, let us work out the case $k=0$
where all $\alpha$'s and $\rho$'s are one. We get a single equation
$f(0,W_1)=-W_1=f(W,W)$, which yields
\eqn\redkzer{\eqalign{
W_1&= W-gW^3 \qquad {\rm (Q.)}\cr 
W_1&=1+g W^3(1-3 g W) \qquad {\rm (E.T.)}\cr}}
in agreement with the results \forWone.

The case $k=1$ of the statistics of nearest neighboring edges and vertices
will be studied extensively in the next section. 

For fixed $k$, we have also access to the statistics of finitely
distant edges and vertices from a vertex in {\it infinite} graphs,
defined as the limit $N\to \infty$ of the statistics at fixed $N$. 
In particular, we may define 
\eqn\gade{\Gamma(\{\alpha_1,\cdots,\alpha_k;\rho_1,\cdots,\rho_k\})=
\lim_{N\to\infty} 
\Gamma_N(\{\alpha_1,\cdots,\alpha_k,1,\cdots;\rho_1,\cdots,\rho_k,1,\cdots\}) }
Eq.\invstat\ translates into
\eqn\invge{
\Gamma(\{\alpha_1,\cdots,\alpha_k;\rho_1,\cdots,\rho_k\})= 
{\int_0^{\alpha_1} {d\alpha \over \alpha}
W_1(\{\alpha,\alpha_2,\cdots,\alpha_k,1,\cdots;\rho_1,\rho_2,\cdots,\rho_k,1,\cdots\})
\vert_{\rm sing}\over
\int_0^1{d\alpha \over \alpha}
W_1(\{\alpha,1,\cdots;1,1,\cdots\})\vert_{\rm sing}}}
where the subscript $\vert_{\rm sing}$ refers to the singular part of the 
corresponding function of $g$ as $g$ approaches the critical value
$g_c=1/12$ (Q.) or $g_c=1/8$ (E.T.), proportional to
$(g_c-g)^{3/2}$ both in the numerator and denominator of eq.\invge, as we
shall see in next section.

\newsec{Nearest neighbor statistics: general solution}

In this section as well as in the next two ones, we address the question of {\it nearest}
neighbor statistics, for which explicit and particularly simple formulas
may be written. Here we concentrate first on the case of quadrangulations and then
repeat our analysis for Eulerian triangulations.  

\subsec{Quadrangulations}

Combining the equations \splitmaster\ and \intmagic\ for $k=1$
in the case of quadrangulations, we obtain the following third
degree algebraic equation, with $\alpha\equiv \alpha_1$ and $\rho\equiv \rho_1$  
\fig{The rooted planar quadrangulations with up to two faces, and the
corresponding weights $\alpha$ per nearest neighboring edge (represented 
in thick solid line) of the origin (labeled $0$), and $\rho$ per nearest neighboring 
vertex (represented in grey, and labeled $1$). These weights match the
expansion (4.2) of $W_1$.}{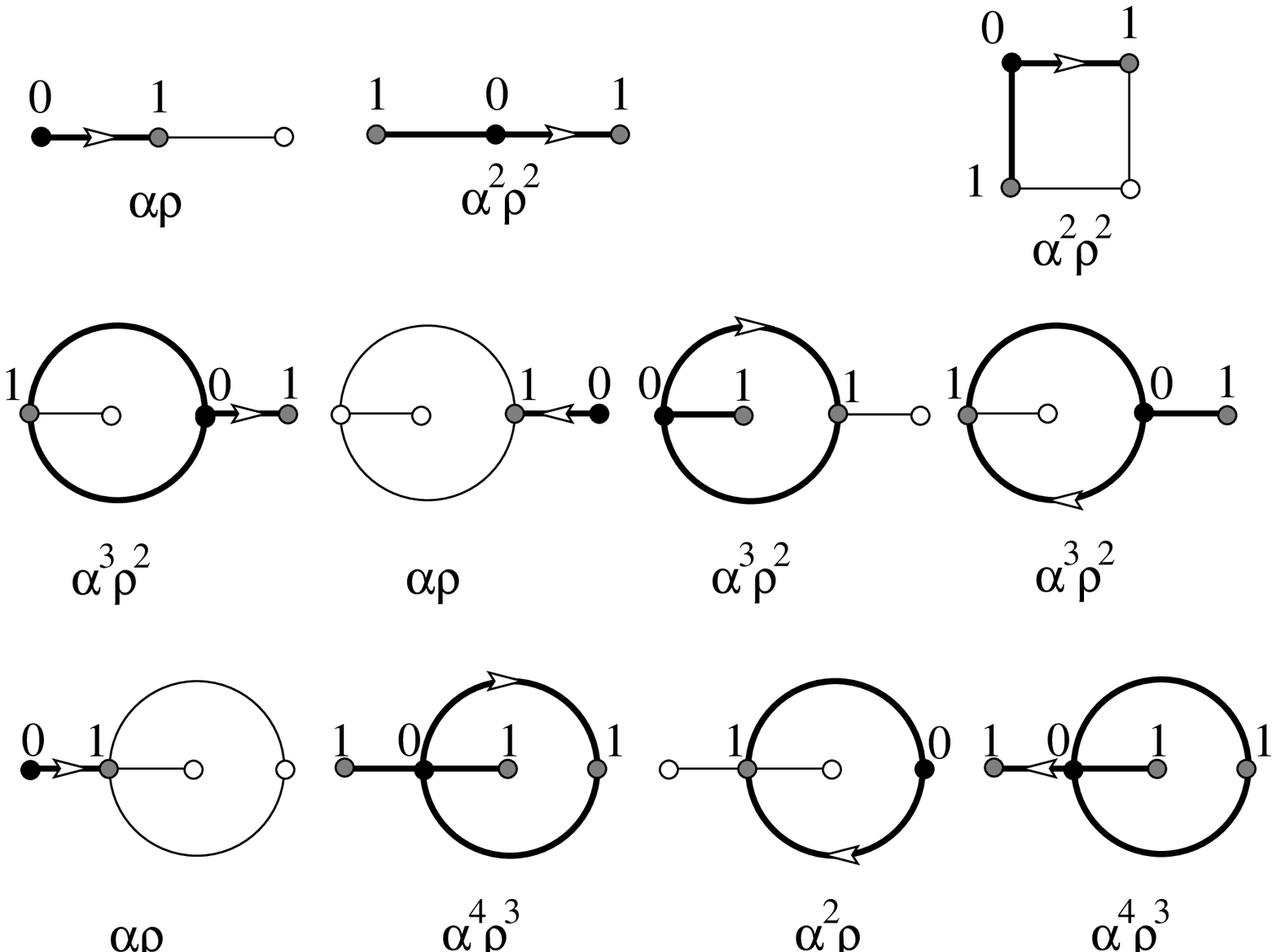}{12.cm}
\figlabel\firfewqua
\eqn\rzerqua{ (W_1-\rho)(W_1(\alpha-1)+\rho-1)
-g \alpha W_1(W_1^2\alpha(\alpha-1)+\alpha\rho W_1+W (W-2))+2 g^2\alpha
W^3 W_1=0}
We must pick the unique solution of eq.\rzerqua\ 
with a $g$ series expansion, and such that $W_1=\rho+O(g)$. We get
\eqn\readqua{W_1=\rho+g \alpha \rho(1+\alpha\rho)
+g^2\alpha\rho(2+\alpha+\alpha\rho+3\alpha^2\rho+2\alpha^3\rho^2)+O(g^3)}
up to order $2$ in $g$, in agreement with the direct enumeration of the
quadrangulations with up to two faces (see Fig.\firfewqua). 

We may now study in detail the statistics of nearest neighbors in {\it infinite}
quadrangulations by letting $g\to g_c=1/12$.
Close to this critical point we write
\eqn\scaqua{ g={1\over 12}(1-\epsilon^2), \qquad W={2\over 1+\epsilon}} 
and substitute the expanded solution 
$W_1=a+b \epsilon+c \epsilon^2+d \epsilon^3+O(\epsilon^4)$
into the cubic equation \rzerqua. At order $0$ in $\epsilon$, we find
that
\eqn\wa{F(a\vert \alpha;\rho)\equiv 36\rho(\rho-1)
+4 a (9-\alpha-18\rho+9\alpha\rho)
+3 a^2(12(1-\alpha)+\alpha^2\rho)+3 a^3\alpha^2(\alpha-1)=0}
which fixes $a$ as the unique solution with value $4/3$ at $\alpha=\rho=1$,
extracted from the solution \forWone\ at $\theta=\sqrt{\epsilon}=0$. 
Alternatively, this solution also satisfies
$a= \rho+O(\alpha)$ as a series of $\alpha$, and this latter
property also fixes it entirely among the three possible solutions of $F(a)=0$. 
Moreover we find at order one in $\epsilon$ that
$b=0$, and the most singular ($\propto (g_c-g)^{3/2}$) term $d$, 
obtained at order $3$ in $\epsilon$, is related to $a$ via
\eqn\dea{ W_1(\alpha;\rho)\vert_{\rm sing}=d={8 \alpha a\over \partial_a F(a\vert \alpha;\rho)} }
with $F$ as in eq.\wa.
Finally, let us perform the change of variable $\alpha\to A=\alpha a$ in the integrals
of eq.\invge. Using the explicit expression of $\alpha$ in terms of $A$:
\eqn\alphaofa{\alpha={3\over 2}A\,{12 A+A^2+24 \rho -12-(2-A)\sqrt{(18-A)(2-A)}
\over 36 \rho (\rho+A-1)+A(3A\rho+3A^2-4)} }
as well as
$d\alpha/dA=-\partial_aF/((A/\alpha)\partial_a F-\alpha\partial_\alpha F)$,
we get
\eqn\newgaeq{\Gamma\equiv \Gamma(\alpha;\rho)= 
{\int_0^A dy {8(y/\alpha)\over (y/\alpha)\partial_a F(y/\alpha|\alpha,\rho)-\alpha
\partial_\alpha F(y/\alpha|\alpha;\rho)}\over
\int_0^{4/3} dy {8y\over (y\partial_a F(y|1;1)
-\partial_\alpha F(y|1;1)}}
={1\over 2}\left(\sqrt{18-A\over 2-A} -3\right)}
This remarkably simple formula is easily inverted into
\eqn\invsimp{ A=2 {\Gamma (\Gamma+3)\over (\Gamma+1)(\Gamma+2)} }
itself mysteriously reminiscent of the form of the solution \results.
To determine $\Gamma$, we finally
note that $a=A/\alpha$ being a root of the cubic equation \wa, we may write 
a cubic equation for $\Gamma$ as well:
\eqn\cubgam{6 \Gamma(1+\Gamma)(3+\Gamma)-\alpha\big(2\Gamma(1+4\Gamma+\Gamma^2)+3\rho 
(1+\Gamma)^2(2+\Gamma)\big)=0}
which fixes $\Gamma$ uniquely by the condition that $\Gamma=1$ for $\alpha=\rho=1$,
or alternatively by $\Gamma=0$ when $\alpha=0$.

\subsec{Eulerian triangulations}

Combining the equations \splitmaster\ and \intmagic\ for $k=1$
in the case of Eulerian triangulations, and taking again $\alpha_1\equiv \alpha$, $\rho_1\equiv
\rho$, we get the following algebraic equation of degree three
for $W_1=Z_1/\alpha$:
\eqn\algtrico{\eqalign{
&(\rho-W_1)(W_1(\alpha-1)+\rho-1)+g \alpha W_1(W(W-2)+(\alpha-1)W_1(W_1+1)\cr
&+\rho(1+2W_1-W_1\alpha-\rho))+g^2 \alpha W_1W^2(1-2W)+g^3\alpha W^4 W_1=0\cr}}
\fig{The rooted planar Eulerian triangulations with up to $6$ faces, and the
corresponding weights $\alpha$ per nearest neighboring edge (represented 
in thick solid line) from the origin (labeled $0$), and $\rho$ per nearest neighboring 
vertex (represented in grey, and labeled $1$). These weights match the
expansion (4.11) of $W_1$.}{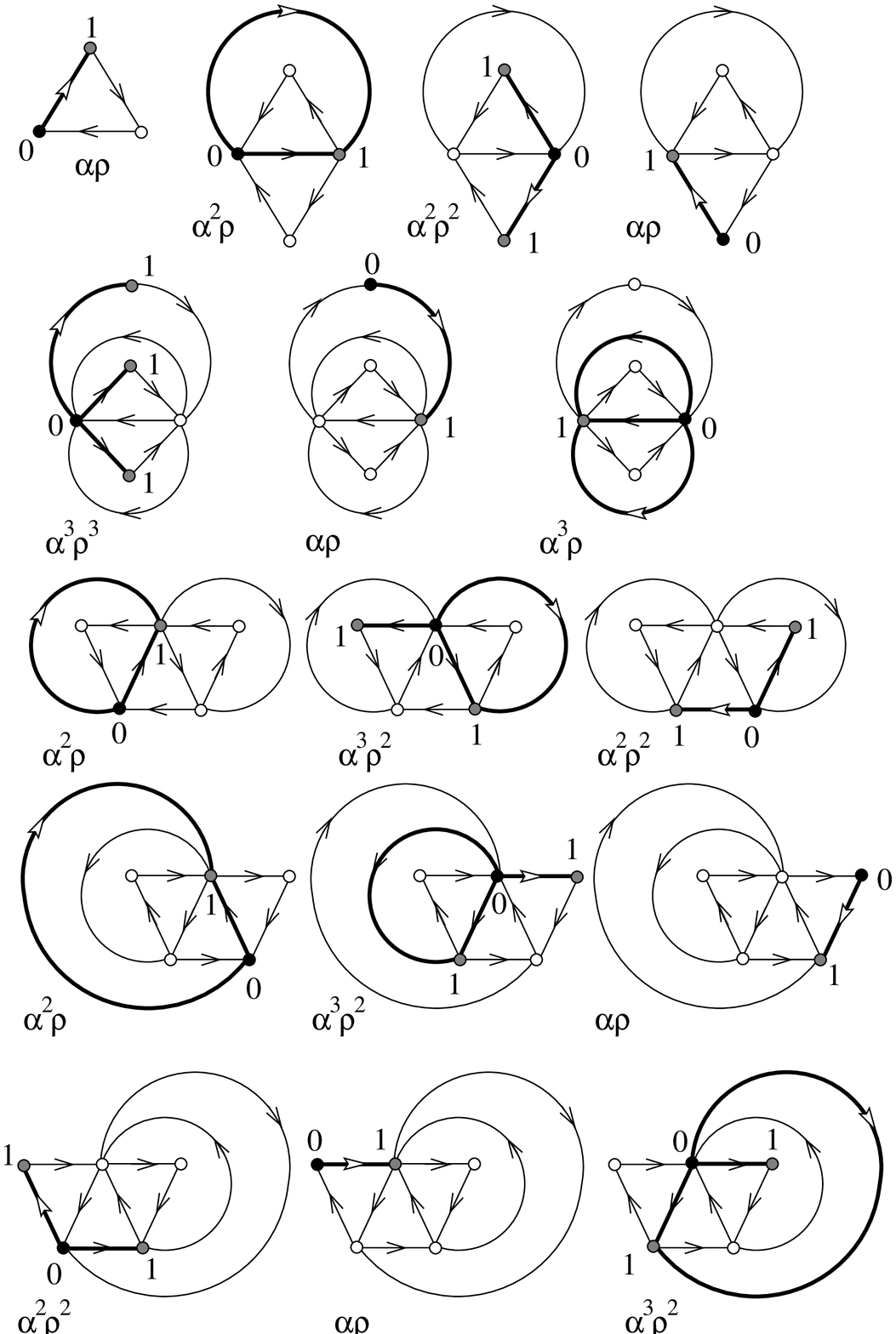}{11.cm}
\figlabel\firfewtri
Again, we must pick the unique solution of eq.\algtrico\ 
with a $g$ series expansion, such that $W_1=\rho+O(g)$, whose first
terms read
\eqn\readtri{W_1=\rho+g\alpha \rho+g^2 \alpha \rho(1+\alpha+\alpha\rho)
+g^3\alpha\rho(3+2\alpha+\alpha^2+2\alpha \rho+3 \alpha^2\rho +\alpha^2\rho^2)+O(g^4)}
up to order $3$ in $g$, in agreement with the graphs of Fig.\firfewtri.

Close to the critical point we write
\eqn\scatri{ g={1\over 8}(1-\epsilon^2), \qquad W={2\over 1+\epsilon}}
and substitute the expanded solution
$W_1=a+b \epsilon+c \epsilon^2+d \epsilon^3+O(\epsilon^4)$
into the cubic equation \algtrico. At order $0$ in $\epsilon$, we find
that
\eqn\wat{\eqalign{
F(a\vert \alpha;\rho)&\equiv 
32\rho(\rho-1)
+a (32-5\alpha-64\rho+4\alpha\rho(9-\rho))\cr
&+4 a^2\big((\alpha-1)(\alpha-8)-\alpha\rho(\alpha-2)\big)+4 a^3\alpha(\alpha-1)=0\cr}}
which fixes $a$ as the unique solution with value $5/4$ at $\alpha=\rho=1$,
extracted from the solution \forWone\ at $\theta=\sqrt{\epsilon}=0$.
As before, this solution also satisfies
$a=\rho+O(\alpha)$, 
which fixes it entirely among the three possible solutions of $F(a)=0$.
Again, we find at order one in $\epsilon$ that
$b=0$, and the most singular ($\propto (g_c-g)^{3/2}$) term $d$
is obtained at order $3$ in $\epsilon$.
It is related to $a$ via
\eqn\deat{W_1(\alpha;\rho)\vert_{\rm sing}=d={8 \alpha a\over \partial_a F(a\vert \alpha;\rho)} }
with $F$ as in eq.\wat.
Finally, let us perform the change of variable $\alpha\to A=\alpha a$ in the integrals
of eq.\invge. Using the explicit expression of $\alpha$ in terms of $A$:
\eqn\alphaofat{\alpha=
2A\,{9A-A^2+16\rho-2A\rho-8-(2-A)\sqrt{(8-A)(2-A)}\over 
32\rho(\rho-1)
+36A\rho-4A^2\rho-4A\rho^2+4A^2-5A } }
as well as
$d\alpha/dA=-\partial_aF/((A/\alpha)\partial_a F-\alpha\partial_\alpha F)$,
we get
\eqn\newgaet{\Gamma\equiv \Gamma(\alpha;\rho)= 
{\int_0^A dy {8(y/\alpha)\over (y/\alpha)\partial_a F(y/\alpha|\alpha,\rho)-\alpha
\partial_\alpha F(y/\alpha|\alpha;\rho)}\over
\int_0^{5/4} dy {8y\over (y\partial_a F(y|1;1)
-\partial_\alpha F(y|1;1)}}
=\sqrt{8-A\over 2-A} -2}
easily inverted into
\eqn\invsimpt{ A=2 {\Gamma (\Gamma+4)\over (\Gamma+1)(\Gamma+3)} }
also reminiscent of the form of the solution \results.
We may again use eq.\wat\ to write a cubic equation for $\Gamma$:
\eqn\cubgamt{4 \Gamma(\Gamma+2)(\Gamma+4)-\alpha(\Gamma+1)\big(\Gamma(\Gamma+5)+2\rho(\Gamma+2)
(\Gamma+3)\big) =0}
and $\Gamma$ is fixed uniquely by the condition that $\Gamma=1$ for $\alpha=\rho=1$,
or alternatively by $\Gamma=0$ when $\alpha=0$.

\newsec{Nearest neighbor statistics: probabilities and moments}

\subsec{Quadrangulations}

We may extract from the solution of eq.\cubgam\ various probabilities and moments for the numbers of
nearest neighboring edges or vertices of a given vertex in infinite quadrangulations. 
Taking for instance $\alpha=1$, the solution reads
\eqn\solaone{ \Gamma(1;\rho)= {2\over \sqrt{4-3\rho}} -1=\sum_{n\geq 1}\rho^n 
\left({3\over 16}\right)^n
{2n \choose n}  }
in which we read the probability
\eqn\probq{ P(n)=\left({3\over 16}\right)^n {2n \choose n} }
that a vertex have $n$ nearest neighboring vertices.
Analogously, taking $\rho=1$, we get
\eqn\rhoneq{ \Gamma(\alpha;1)= {1\over 2} \left( \sqrt{3(2+\alpha)\over 6-5\alpha}-1\right)
={1\over 3}\alpha+{1\over 6}\alpha^2+ {13\over 108} \alpha^3+\cdots }
in which we read the probabilities $1/3,1/6,13/108,\cdots$ for 
having $1,2,3,\cdots$ nearest neighboring edges.

The formulae \solaone\ and \rhoneq\ also yield the various moments of the distribution
of the numbers $n_1$ and $m_1$ of nearest neighboring vertices and edges, upon writing
$\rho=e^u$ (resp. $\alpha=e^v$) and expanding around $u=0$ (resp. $v=0$).
For instance we get
\eqn\instmom{\eqalign{ &\langle n_1\rangle= 3, \quad \langle n_1^2\rangle={33\over 2},
\quad \langle n_1^3\rangle={579\over 4} \cr  
&\langle m_1\rangle= 4, \quad \langle m_1^2\rangle={100\over 3},
\quad \langle m_1^3\rangle={1372\over 3} \cr}}
The value $\langle m_1\rangle= 4$ does not come as a surprise, as for each quadrangulation
$\cal G$ with $N$ faces, we have $\sum_{v\in {\cal G}} m_1({\cal G},v) =4 N$ (twice the
total number of edges), while there are $N+2$ vertices, hence $\langle m_1\rangle_N=4N/(N+2)\to 4$
for large $N$. 

More generally, we have access to the joint probabilities for the numbers of nearest neighboring
edges and vertices and to the associated mixed moments by expanding $\Gamma$ as a double
power series of $\alpha$ and $\rho$ around $0$ (probabilities) or $1$ (moments). 

Finally, we also have access to the conditional probabilities of having $n$ nearest neighboring
vertices {\it given that there are no multiple nearest neighbors}. This is achieved by
taking $\rho=t/\alpha$, in which case $\Gamma=\langle \alpha^{m_1-n_1}t^{n_1}\rangle$, 
and then sending $\alpha\to 0$ to retain only the configurations where $m_1=n_1$ (as $m_1\geq n_1$
always). This results in the generating function $\Pi(t)/\Pi(1)$ for the above conditional 
probabilities, where
\eqn\newga{\Pi(t)\equiv \lim_{\alpha\to 0} \Gamma(\alpha;{t\over \alpha})=
{1\over 2}\left(\sqrt{18-t\over 2-t}-3\right)= {1\over 3}t+{7\over 54}t^2+{53\over 972}t^3+\cdots}
and in particular we get the probability of having no multiple
nearest neighbor $\Pi(1)=(\sqrt{17}-3)/2$.
We note that from eq.\newgaeq\ we have $\Pi(A)=\Gamma$. This relation is not surprising, once
understood 
in the language of well-labeled trees. 
Indeed, the trees contributing to $\Pi(t)$ are those associated with graphs without multiple
nearest neighbors of the origin, i.e. those in which the vertices labeled $1$ are all
terminal (univalent) vertices and receive a weight $t$. Replacing this weight $t$ by 
$Z_1=\alpha W_1$ reconstructs the full generating function contributing to $\Gamma$.
For large graphs, this corresponds to the substitution $t\to A$ in $\Pi$.

\subsec{Eulerian triangulations}

Repeating the various specializations of previous section of the solution to eq.\cubgamt,
we get for $\alpha=1$:
\eqn\alont{ \Gamma(1;\rho)= {1\over 2}\left(\sqrt{27-2\rho\over 3-2\rho}-3\right)={4\over 9}\rho+
{56\over 243}\rho^2+{848\over 6561}\rho^3+\cdots}
leading to the probabilities $4/9,56/243,848/6561,\cdots$ 
that a vertex have $1,2,3,\cdots$ nearest neighboring vertices.
For $\rho=1$, we get:
\eqn\rhoneqt{ \Gamma(\alpha;1)={2\over \sqrt{4-3\alpha}} -1}
in which we read the probability
\eqn\probqt{ P(m)=\left({3\over 16}\right)^m {2m \choose m} }
for having $m$ nearest neighboring edges.
Remarkably, these probabilities for nearest neighboring {\it edges} in Eulerian triangulations
coincide with those of eq.\probq\ for nearest neighboring {\it vertices} in
quadrangulations. Note also that the probability \probqt\ is very similar to that
obtained in Ref.\GKY\ for ordinary triangulations with no multiple edge and no loop. 

The corresponding moments read
\eqn\momentri{\eqalign{ &\langle n_1\rangle= {12\over 5}, 
\quad \langle n_1^2\rangle={1212\over 125},
\quad \langle n_1^3\rangle={190236\over 3125} \cr
&\langle m_1\rangle= 3, \quad \langle m_1^2\rangle={33\over 2},
\quad \langle m_1^3\rangle={579\over 4} \cr}}
The value $\langle m_1\rangle=3$ can be obtained by explicitly computing $\langle
m_1\rangle_N=3N/(N+2)\to 3$ for large $N$.

Finally, we also have the conditional probabilities of having $n$ nearest neighboring
vertices given that there are no multiple nearest neighbors. Repeating the limiting process of
previous section we arrive at
\eqn\newgat{ \Pi(t)\equiv \lim_{\alpha\to 0} \Gamma(\alpha;{t\over \alpha})=
\sqrt{8-t\over 2-t} -2={3\over 8}t+{39\over 256} t^2+{267\over 4096}t^3+\cdots}
and in particular we get the probability of having no multiple
nearest neighbor $\Pi(1)=\sqrt{7}-2$. 
As before, we have $\Pi(A)=\Gamma$ of eq.\newgaet.

\newsec{Heuristic interpretation: Phase diagram}

The results of previous section display a number of critical
values of $\alpha$ and $\rho$ at which the various generating functions
become singular. For quadrangulations we have singular points
at $(\alpha=1;\rho=4/3)$ and $(\alpha=6/5;\rho=1)$, easily read off
eqs.\solaone\ and \rhoneq. The value $\rho^{-1}=3/4$ (resp. $\alpha^{-1}=5/6$)
governs the exponential decay
of the corresponding probabilities, behaving as $(3/4)^n$ (resp. as $(5/6)^m$).  
Similarly, for Eulerian triangulations, we have the particular points
$(\alpha=1;\rho=3/2)$ and $(\alpha=4/3;\rho=1)$, read off eqs.\alont\ and \rhoneqt.
These points are particular points on a more general critical curve in the 
$(\alpha,\rho)$ plane, whose location may be easily obtained from the following heuristic
argument.

Labeled trees
may be thought of as 
trees embedded in a one-dimensional target space,
by viewing the labels as the positions of the vertices in the target space. 
In this language, the generating function $W_j$ corresponds to trees with 
a root at {\it position} $j$.
Without the constraint that the labels remain positive, all $W_j$'s
are equal to $W$ of eq.\formW, with $W\vert_{g^N}=3^N c_N$ (Q.) or
$2^N c_N$ (E.T.), counting arbitrary rooted planar trees with $N$ edges
(in number $c_N$) together with $3^N$ (resp. $2^N$) embeddings accounting for the $3$
(resp. $2$) possibilities of labeling a vertex according to  
the label of its immediate ascendent vertex.
\fig{Phase diagram in the $(\rho,\alpha)$ plane for rooted trees embedded in
a one-dimensional target space with a wall and weights $\alpha$ (resp. $\rho$) per
edge (resp. vertex) touching the wall. A critical curve separates a region where
the tree is attracted by the wall from a region where it is repulsed by it.
This curve possesses a higher order (tricritical) point represented by a black dot.
The corresponding leading singular behavior of $W_1$ is indicated. 
The indicated values of $\alpha$ and $\rho$ refer to well-labeled trees associated 
to quadrangulations (Eulerian triangulations).}{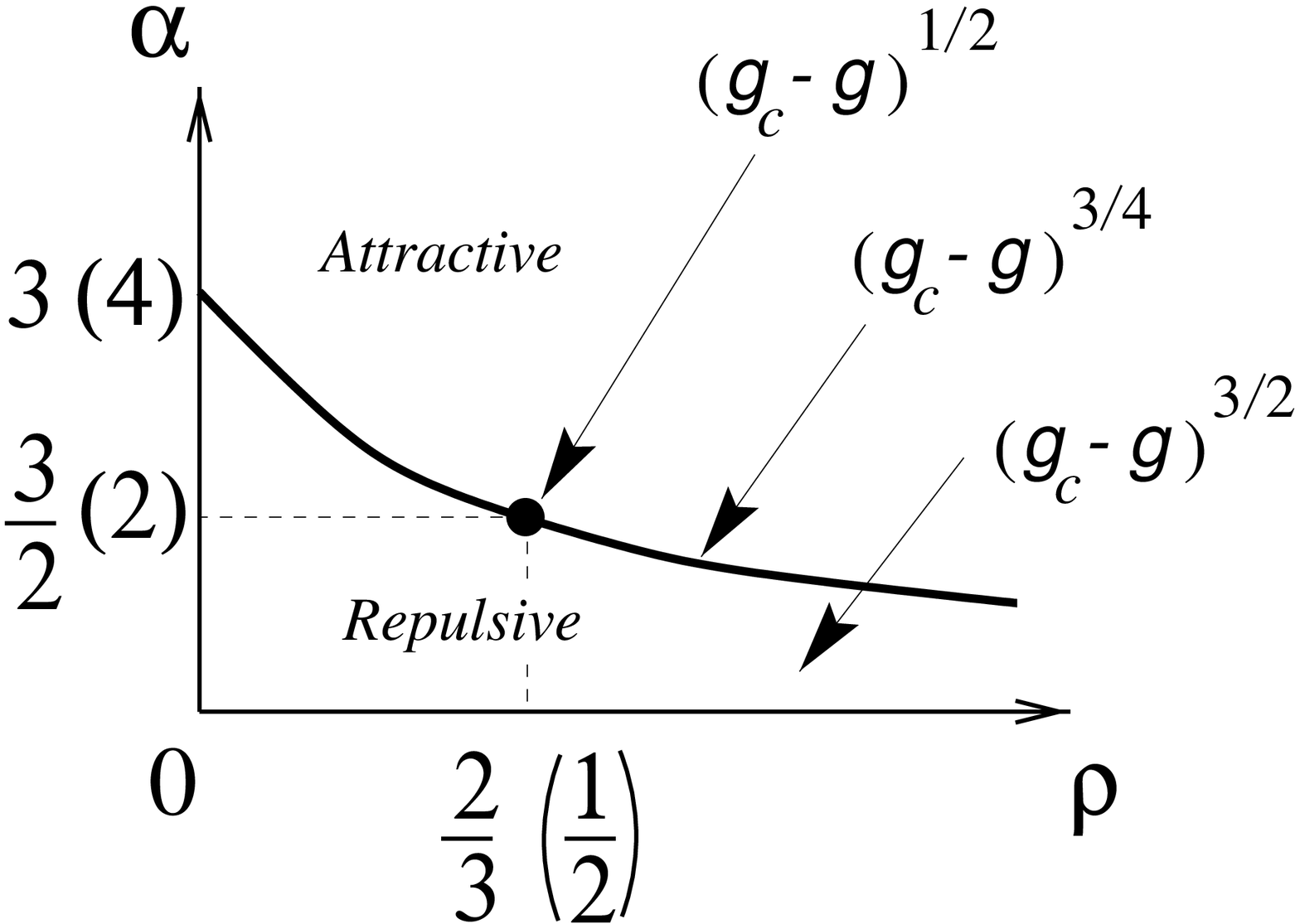}{8.cm}
\figlabel\phasedia
The positivity of all labels simply restricts the target space
to a half-line, by introducing a hard wall at the origin, whose effect is to
reduce the entropy when compared with the no-wall situation.  
On the other hand, the weights $\alpha$ and $\rho$ of Sect.4 act as energy weights
when an edge or a vertex touches the wall,  
which may help restore the balance between the reduction of entropy and a possible
gain in energy. The critical curve corresponds to an
unbinding transition at which the balance
is realized. Its location may be obtained as follows.
For each vertex of the tree reaching position $1$,
we have an energy gain $\rho \alpha^{k}$, where $k$ is the number of nearest neighbors
of the vertex on the tree. On the other hand, the entropy loss is
$(2/3)^{k-1}$ (Q.) or $1/2^{k-1}$ (E.T.) as the $k-1$ descendents of this vertex
have lost one of their $3$ (resp. 2) possible positions.  
In an infinite rooted planar tree, the probability of having $k$ neighbors
is $1/2^k$, hence the balance is reached when
\eqn\balance{\eqalign{1 &= \sum_{k\geq 1} {1\over 2^k} \rho \alpha^k
\left({2\over 3}\right)^{k-1}={3\over 2} {\rho \alpha\over 3-\alpha} 
\qquad {\rm (Q.)}\cr
1 &= \sum_{k\geq 1} {1\over 2^k} \rho \alpha^k
{1\over 2^{k-1}}= {2\rho \alpha\over 4-\alpha}
\qquad {\rm (E.T.)}\cr}}
This displays the critical lines $2(3-\alpha)=3\rho\alpha$ (Q.) or $4-\alpha=2\rho\alpha$
(E.T.) separating a phase in which the tree is ``attracted" by the wall from a phase
where it is ``repulsed" by it (see Fig.\phasedia). 
All pairs $(\alpha;\rho)$ mentioned above belong to these critical lines.
A special point on these lines corresponds to the
{\it exact balance} for {\it each} separate value of the valence $k$, namely when
$\rho \alpha^k=(3/2)^{k-1}$ (Q.) or $\rho \alpha^k=2^{k-1}$ (E.T.) for all $k\geq 1$,
i.e. $\alpha=3/2,\rho=2/3$ or $\alpha=2,\rho=1/2$ (E.T.). At this special
point, the effect of the wall is exactly suppressed and the half-line solution 
coincides with the no-wall solution, namely $Z_j=W$ for all $j\geq 1$. 
Therefore at this special point, the
singularity of $W_1=\rho W$ is proportional to $(g_c-g)^{1/2}$, instead of the generic
singularity $\propto (g_c-g)^{3/2}$ already observed away from the critical line.
It is easily checked that the generic singularity
along the critical line is of the form $(g_c-g)^{3/4}$, easily obtained by expanding
the equations \rzerqua\ and \algtrico\ now in powers of $\theta=\sqrt{\epsilon}$
and solving order by order for the coefficients of $W_1=A+B\theta+C\theta^2+D \theta^3+\cdots$
These various critical behaviors translate into different asymptotics for the 
weighted sums over graphs of size $N$, 
of the form $g_c^{-N}/N^{\gamma}$ with $\gamma=5/2$ (most generic case,
away from the critical line), $\gamma=7/4$ (generic on the critical line) and
$\gamma=3/2$ (special point on the critical line). 

Beside the above heuristic argument, the critical lines may also be derived by imposing
that both $F(a\vert\alpha;\rho)=0$ and $\partial_a F(a\vert\alpha;\rho)=0$
(with $F$ as in eqs.\wa\ or \wat), while the
special (tricritical) point corresponds to having in addition 
$\partial_a^2 F(a\vert\alpha;\rho)=0$.

\newsec{Generalization to more distant neighbors}

\subsec{General scheme for getting $\Gamma$}

The results of Sects. 4 and 5 may be generalized so as to include 
more distant neighbors, say up to $k$-th neighboring edges and vertices.
This involves explicitly solving
the equations \splitmaster\ and \intmagic. The resulting algebraic
equation for $W_1$ however grows in degree with $k$, and one still
has to integrate once with respect to $\alpha_1$ to finally
obtain the statistical average $\Gamma$ of eq.\invge.  
Nevertheless, we have found a particularly simple alternative way of getting
$\Gamma$ by relating it to a single quantity $t$ parametrizing
all $Z_j$'s {\it at the critical point}. 
More precisely, when $g=g_c$, with $g_c=1/12$ (Q.) or $g_c=1/8$ (E.T.) 
and $W=W_c=2$, $x=x_c=1$ (as readily seen from eqs.\formW\ and \formu), 
the convergence
condition \intmagic\ is trivialized by a change of variables
on $Z_k=Z_k^c$ borrowed from the exact solution. Namely, if we write
\eqn\chvars{Z_j^c\equiv\zeta_j(t_j), \ \ {\rm with}\ \ 
\left\{\matrix{\zeta_j(t)=2{(t+j)(t+j+3)\over (t+j+1)(t+j+2)} & {\rm (Q.)}\cr
\zeta_j(t)=2{(t+j)(t+j+4)\over (t+j+1)(t+j+3)} & {\rm (E.T.)}\cr}\right.}
for all $j\geq k$, then the condition $f(Z_j^c,Z_{j+1}^c)=f(2,2)$ amounts
to $t_j=t_{j+1}$, hence $t_j=t$ for all $j\geq k$. 
The value of $t$ is obtained via the set of $k$ equations \splitmaster, which we
feed with the two ``final conditions" $Z_k^c=\zeta_k(t)$ and $Z_{k+1}^c=
\zeta_{k+1}(t)$,
allowing to express backward all $Z_j^c$'s with $j<k$ as rational functions of $t$
and to finally get one algebraic equation for $t$ with $\alpha$'s and
$\rho$'s as parameters. The proper solution is then uniquely characterized 
by the fact that when all $\alpha$'s and $\rho$'s are one, we have $t=0$, as 
can be read
off the solution \results\ at criticality, where $W=2$ and $u_j\propto j$.

Quite magically, it turns out that $t$ {\it may be identified with 
the desired function} $\Gamma$,
up to a translation independent of $\alpha_1$.
More precisely, we have
\eqn\gaT{ 
\Gamma(\{\alpha_i;\rho_i\})=t(\{\alpha_1,\alpha_2,\cdots;\rho_1,\rho_2,\cdots\})-
t(\{0,\alpha_2,\cdots;\rho_1,\rho_2,\cdots\}) }
as $\Gamma$ must vanish when $\alpha_1=0$ (note that when $\alpha_1=0$,
$Z_1=0$ and $t$ does not depend on $\rho_1$ either). 
This remarkable identity may be checked at $k=1$ where 
$t(\{0,1,1,\cdots;\rho_1,1,1,\cdots\})=t(\{0,1,1,\cdots;0,1,1,\cdots\})=-1$ 
(the solution is the same as when all $\alpha$'s
and $\rho$'s are one except for a shift in the indices) and the 
relation $\Gamma=t+1$ is read off eqs. \invsimp\ and \invsimpt\ with
$A=Z_1^c$. 
For arbitrary $k$, the identification \gaT\ will be proved in the
next section thanks to scaling arguments. 

The value of $\Gamma$ may therefore be extracted as follows:
let $G(t)\equiv G(t\vert\{\alpha_i;\rho_i\})=0$ denote the algebraic
equation satisfied by $t$. Then if we take $\alpha_1=0$
we get an algebraic equation $G_0(t_0)=0$ for the translation parameter
$t_0\equiv t(\{0,\alpha_2,\cdots;\rho_1,\rho_2,\cdots\})$ of eq.\gaT. 
We therefore simply have to eliminate $t_0$ between the two equations
$G(\Gamma+t_0)=0$ and $G_0(t_0)=0$ to get an equation for $\Gamma$.
This equation might be sometimes factorized into smaller degree equations, 
but the initial condition $\Gamma=O(\alpha_1)$ always
selects only one of them. For illustration, we will recover in Sect. 7.3 below
the results for $\Gamma$ at $k=1$, and further present the solution
for $k=2$. 

\subsec{The reason why $t=\Gamma+t_0$}

In this section we show that the parameter $t$ occurring in the changes of 
variables
\eqn\chvarsapp{\eqalign{ Z_j^c=2{(t+j)(t+j+3)\over (t+j+1)(t+j+2)} \qquad {\rm (Q.)}\cr
Z_j^c=2{(t+j)(t+j+4)\over (t+j+1)(t+j+3)} \qquad {\rm (E.T.)}\cr}}
coincides with the statistical average $\Gamma$ up to an additive constant
independent of $\alpha_1$.
We simply have to show that $dt/d\alpha_1=d\Gamma/d\alpha_1$.
Let us first find a simpler characterization of $t$, by considering
$Z_j^c$ at large $j$.
Expanding $Z_j^c$ at large $j\geq k$, we find 
\eqn\expanZ{\eqalign{ Z_j^c=2- {4\over j^2}+{8\over j^3}(t+{3\over 2})
+O({1\over j^4})
\qquad {\rm (Q.)} \cr
Z_j^c=2- {6\over j^2}+{12\over j^3}(t+2)+O({1\over j^4})
\qquad {\rm (E.T.)} \cr}}
therefore 
$t$ enters only the subleading correction $\propto 1/j^3$ of $Z_j^c$. 
Let us now go back 
to the combinatorial interpretation
of $Z_j$ (for large enough $j$) as a series of $g$, before taking $g\to g_c$. 
We see that $dZ_j/d\alpha_1$ is 
the generating function for well-labeled 
trees with a root at position $j$ and a {\it marked oriented edge} originating 
from 
a vertex at position $1$. In terms of graphs\foot{Here and throughout the 
appendix, the
mention graph refers to either quadrangulations or Eulerian triangulations.}, 
consider the generating function $G_j(\{\alpha_i;\rho_i\})$ 
for graphs with a marked vertex
(origin) and a marked edge from a vertex at geodesic distance $j-1$ from the 
origin
to a vertex at geodesic distance $j$ from the origin. Then we have
\eqn\wehazg{\eqalign{ 
G_j&(\{\alpha_1,\alpha_2,\cdots,\alpha_k;\rho_1,\rho_2,\cdots,\rho_k\})\cr
&=Z_j(\{\alpha_1,\alpha_2,\cdots,\alpha_k;\rho_1,\rho_2,\cdots,\rho_k\})
-Z_{j-1}(\{\alpha_2,\alpha_3,\cdots,\alpha_k;\rho_2,\rho_3,\cdots,\rho_k\})\cr}}
Indeed, in the tree language, $G_j$ generates the well-labeled trees with root 
at position
$j$ and which {\it do reach} the position $1$, therefore expressed via \wehazg\ 
by
subtracting from the generating function $Z_j(\{\alpha_i;\rho_i\})$
the contribution of trees which do not reach position $1$, also 
equal
to that of trees with root at position $j-1$, provided we shift the whole 
picture
by $-1$, thus removing the terms $\alpha_1,\rho_1$ pertaining to position $1$. 
Taking a derivative of eq.\wehazg\ with respect to $\alpha_1$, we see that 
$dZ_j/d\alpha_1= dG_j/d\alpha_1$. 
With this identification, we have
\eqn\wehavee{{\alpha_1 {d Z_j\over d\alpha_1}|_{g^N}\over W_1(\{1;1\})|_{g^N}}
={\langle m_1 m_j \prod \alpha_i^{m_i}\rho_i^{n_i}\rangle_N\over 
\langle m_1\rangle_N }}
When $N\to \infty$, we deduce that
\eqn\wededu{{\alpha_1 {d Z_j\over d\alpha_1}|_{\rm sing}\over 
W_1(\{1;1\})|_{\rm sing}} 
={\langle m_1 m_j \prod \alpha_i^{m_i}\rho_i^{n_i}\rangle\over 
\langle m_1\rangle }}
where $\langle\cdots\rangle$ with no index $N$ refers to the large $N$
limit. For large $j$ we expect the average in the numerator to factor
into $\langle m_j\rangle\langle m_1 \prod \alpha_i^{m_i}\rho_i^{n_i}\rangle$
as the weights are localized around the origin ($k$ is kept finite).
We arrive at the relation
\eqn\relazga{\lim_{j\to \infty}{{d Z_j\over d\alpha_1}|_{\rm sing}\over
\langle m_j\rangle}={W_1(\{1;1\})|_{\rm sing}\over \langle m_1\rangle}
{d\Gamma\over d\alpha_1}={2\over 3}{d\Gamma\over d\alpha_1}}
for both quadrangulations and Eulerian triangulations: 
indeed we have used $\langle m_1\rangle=4$ and $W_1(\{1;1\})|_{\rm sing}=8/3$
(Q.) while $\langle m_1\rangle=3$ and $W_1(\{1;1\})|_{\rm sing}=2$ (E.T.)
from eqs.\dea\ and \deat\ at $\alpha=\rho=1$.

Both $\langle m_j\rangle$ and $d Z_j/d\alpha_1|_{\rm sing}$ may be computed
by considering the usual scaling limit reached by sending 
{\it simultaneously} $g\to g_c$ and $j\to \infty$, say by setting
\eqn\setsca{ g=g_c(1-\theta^2), \qquad j={r\over \theta} }
and letting $\theta\to 0$ (here $\theta^{-1}$ plays the role of a 
correlation length for the models at hand). In this limit,
we have
\eqn\zjexp{ Z_j= 2(1-\theta^2 {\cal U}(r)-\theta^3 {\cal V}(r) -\cdots )}
Upon substituting this expression for $Z_j$ into eqs.\mastersplit, we obtain
at order $4$ and $5$ in $\theta$ the following differential equations
\eqn\difequv{\eqalign{ {\cal U}''&=3\left( {\cal U}^2-1\right) 
\qquad {\cal V}''=6\, {\cal U}\, {\cal V}\qquad {\rm (Q.)} \cr
{\cal U}''&=2\left( {\cal U}^2-1\right) 
\qquad {\cal V}''=4\, {\cal U}\, {\cal V}\qquad {\rm (E.T.)} \cr}}
The behavior of the scaling functions ${\cal U}$ and ${\cal V}$
when $r\to 0$ corresponds to taking $g=g_c$ at fixed $j$, hence we read
off the expansions \expanZ\ of $Z_j^c$ the following leading behaviors
${\cal U}(r)\sim 2/r^2$ and ${\cal V}(r)=-4(t+{3\over 2})/r^3$ (Q.)
while ${\cal U}(r)\sim 3/r^2$ and ${\cal V}(r)=-6(t+2)/r^3$ (E.T.).
The functions ${\cal U}$ and $\cal V$ are further fixed by noting that 
$Z_j\to W=2/(1+\theta^2)=2(1-\theta^2+O(\theta^4))$ at large $j$,
hence ${\cal U}(r)\to 1$ and ${\cal V}(r)\to 0$ when $r\to \infty$.
This results in \GEOD\
\eqn\resforU{\eqalign{ {\cal U}(r)&= {3 \over \sinh^2\left(\sqrt{3\over2}r\right)}
+1 \qquad {\rm (Q.)} \cr
{\cal U}(r)&= {3 \over \sinh^2(r)}
+1 \qquad {\rm (E.T.)} \cr}}
while 
\eqn\fixV{\eqalign{ {\cal V}(r)&=\left(t+{3\over 2}\right){\cal U}'(r) 
\qquad {\rm (Q.)}  \cr
{\cal V}(r)&=(t+2)\, {\cal U}'(r)\qquad {\rm (E.T.)} \cr }}
Expanding eqs.\resforU\ and \fixV\ above as power series of $r$, we have
access to the $\theta$ expansion of $Z_j$ of eq.\zjexp\ by writing
$r=j\theta$, namely
\eqn\expjg{\eqalign{
Z_j =2-\left({4\over j^2}-{8\over j^3}\left( t+{3\over 2}\right)
+O\left({1\over j^4}\right)\right)
&- \theta^4 \left({3\over 5}j^2+{6\over 5}j\left(t+{3\over 2}\right)
+O(1) \right)\cr
&+ \theta^6 \left( {1\over 7} j^4 +{4\over 7}j^3\left(t+{3\over 2}\right)
+O(j^2)\right)+\cdots \cr
Z_j =2-\left({6\over j^2}-{12\over j^3}(t+2)+O\left({1\over j^4}\right)\right)
&- \theta^4 \left({2\over 5}j^2+{4\over 5}j(t+2) +O(1) \right)\cr
&+ \theta^6 \left( {4\over 63} j^4 +{16\over 63}j^3(t+2)+O(j^2)\right)+\cdots \cr}}   
{}From the $\theta^6=(1-g/g_c)^{3/2}$ terms, we read off the singular parts
of $Z_j$ and $dZ_j/d\alpha_1$
\eqn\singzdz{\eqalign{Z_j|_{\rm sing}&={j^4\over 7}+O(j^3)\qquad
{d Z_j\over d\alpha_1}|_{\rm sing}={4 j^3\over 7}{dt\over d\alpha_1}
+O(j^2)
\qquad {\rm (Q.)}\cr Z_j|_{\rm sing}&={4 j^4\over 63}+O(j^3)\qquad
{d Z_j\over d\alpha_1}|_{\rm sing}={16 j^3\over 63}{dt\over d\alpha_1}
+O(j^2)
\qquad {\rm (E.T.)}\cr }}
In particular, the leading contribution to $Z_j|_{\rm sing}$ does
not depend on the $\alpha$'s and $\rho$'s.
To evaluate $\langle m_j\rangle$ at large $j$, we now note that when
all $\alpha$'s and $\rho$'s are one, we have 
\eqn\zjsing{{Z_j(\{1;1\})|_{\rm sing}\over W_1(\{1;1\})|_{\rm sing}}=
\sum_{l=1}^{j}{\langle m_l\rangle\over \langle m_1\rangle}\sim
\left\{\matrix{&{3\over 56} j^4 &{\rm (Q.)}\cr
& & \cr
&{2\over 63} j^4 &{\rm (E.T.)}
\cr}\right.}
where we have used again the values of $W_1(\{1;1\})|_{\rm sing}$ 
above.
The above result for quadrangulations is in agreement with that of
Ref.\GEOD\ (eq.(4.19)).
Using the above values of $\langle m_1\rangle$, we therefore obtain
by differentiating with respect to $j$: 
\eqn\avmj{\langle m_j\rangle \sim \left\{\matrix{&{6\over 7} j^3&
{\rm (Q.)}\cr 
& & \cr
&{8\over 21} j^3&{\rm (E.T.)}\cr}\right.}
at large $j$. Gathering eqs. \relazga, \singzdz\ and \avmj, we finally deduce
that $d\Gamma/d\alpha_1=dt/d\alpha_1$ as announced.

\subsec{Applications}

As a first application of the above scheme, let us re-derive the 
results of Sections 4.1 and 4.2 for $k=1$ with $\alpha_1\equiv\alpha$
and $\rho_1\equiv\rho$. Eqs.\splitmaster\ reduce
to $Z_1(1-g\alpha (Z_1+Z_2))=\rho\alpha$ (Q.) and 
$Z_1(1-g\alpha Z_2)=\rho\alpha$ (E.T.). Substituting 
$Z_1=2(t+1)(t+4)/((t+2)(t+3))$, $Z_2=2(t+2)(t+5)/((t+3)(t+4))$ and 
$g=1/12$ (Q.) or $Z_1=2(t+1)(t+5)/((t+2)(t+4))$, 
$Z_2=2(t+2)(t+6)/((t+3)(t+5))$ and $g=1/8$ (E.T.), we get the equations
for $t$:
\eqn\eqfort{\eqalign{6(t+1)(t+2)(t+4)-\alpha (2(t+1)(t^2+6t+6)+
3\rho(t+2)^2(t+3))& =0 \qquad {\rm (Q.)}\cr
4(t+1)(t+3)(t+5)-\alpha (t+2)((t+1)(t+6)+2 \rho (t+3)(t+4))& =0 
\qquad {\rm (E.T.)}\cr}}
with $t_0=-1$ in both cases at $\alpha=0$. We finally substitute 
$t=\Gamma-1$ in eqs.\eqfort\ to recover eqs. \cubgam\ and \cubgamt.

To conclude this section, 
let us now display the algebraic equations
as obtained via the above scheme 
for $\Gamma$ in the case $k=2$, 
with $\alpha_1=\alpha$, $\alpha_2=1$, 
$\rho_1=\rho$ and $\rho_2=\sigma$.
They read respectively for quadrangulations:
\eqn\lutfin{\eqalign{&6 u^3 \Gamma(u\Gamma+4)(u\Gamma-u+2)(u\Gamma+u+2)
(u\Gamma+2u+2)\cr
&-\alpha \Big( u^3\Gamma(u\Gamma+4)\big(8+4u(2+3\Gamma)
+2u^2(3\Gamma^2+4\Gamma-3)+u^3(\Gamma^3+2\Gamma^2+5\Gamma-4)\cr
&+2u^4(\Gamma+1)(3\Gamma+1)+u^5\Gamma(\Gamma+1)^2\big)
+3\rho(u\Gamma+2)(u\Gamma-u+2)^2(u\Gamma+2u+2)^2 \Big)=0\cr}}
with $u=\sqrt{4-3\sigma}$, and for Eulerian triangulations:
\eqn\lutultim{\eqalign{&192 \Gamma(\Gamma+u)(2\Gamma+u-1)(2\Gamma+u+3)
\cr
&-\alpha(2\Gamma+u-3)(2\Gamma+u+5)\Big( 48\Gamma(\Gamma+u)
+\rho (u^2-1)(2\Gamma+u+1)(2\Gamma+u-1)\Big)=0\cr}}
with $u=\sqrt{(27-2\sigma)/(3-2\sigma)}$.
Again these equations are supplemented by the condition that
$\Gamma=1$ when $\alpha=\rho=\sigma=1$ or alternatively that 
$\Gamma=0$ when $\alpha=0$.

\fig{The probabilities $P(n)$ of having $n$ nearest (resp. next nearest)
neighboring vertices in infinite quadrangulation (a) and
Eulerian triangulations (b), for $n=1,2,...,20$. We note a maximum probability
reached in the case of next nearest neighbors at $n=3$ for quadrangulations and $n=4$ for
Eulerian triangulations, while the probabilities for nearest neighbors 
are strictly decreasing.}{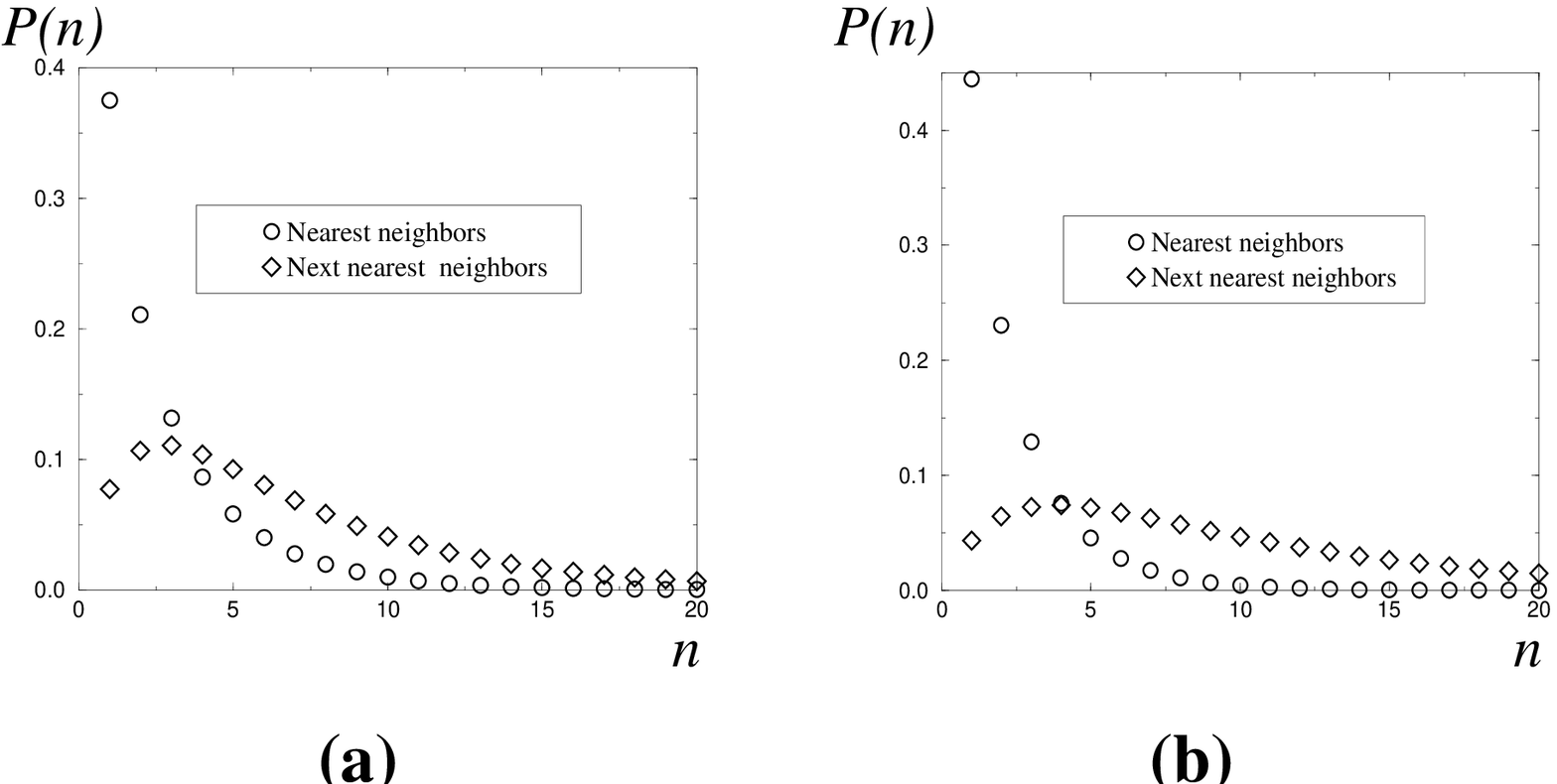}{14.cm}
\figlabel\probas

These equations allow to obtain various moments and probabilities involving 
next nearest neighboring 
vertices in infinite quadrangulations or Eulerian triangulations. 
For instance, we obtain the following first few moments
of the numbers of next nearest neighboring vertices of the origin:  
\eqn\secondmom{\eqalign{\langle n_2\rangle& ={54\over 5},\quad
\langle n_2^2\rangle={24948\over 125},
\quad\langle n_2^3\rangle={17067888\over 3125}\qquad {\rm (Q.)} \cr
\langle n_2\rangle& ={36\over 5},\quad
\langle n_2^2\rangle={2148\over 25},
\quad\langle n_2^3\rangle={4668948 \over 3125}\qquad {\rm (E.T.)} \cr}}
The probabilities $P(n)$ of having $n$ next nearest 
neighboring vertices are displayed in Fig.\probas\ (a) for quadrangulations and (b)
for Eulerian triangulations, and compared with the corresponding probabilities
for nearest neighbors as given by eqs.\probq\ and \alont.

\newsec{Conclusion}

A completely open question regards the generalization to other graphs
than quadrangulations and Eulerian triangulations.
In Ref.\GEOD, we have investigated a number of possible generalizations,
either by allowing for faces with higher valences or by considering
colored graphs, all of which turn out to have generating functions
governed by discrete ``integrable" recursion relations on the 
(suitably defined) 
geodesic distance, and for which compact ``soliton-like"
expressions also exist. These were derived by use of a somewhat different
class of bijections
between the duals of these graphs and some classes of so-called blossom trees with 
{\it constrained} valences [\xref\SCH,\xref\GEOD]. 
To imitate the approach of the present paper, one
would first need to derive a bijection involving geodesic distances, and
with appropriately labeled trees with arbitrary
valences, a task not performed yet. Another more promising approach
would consist in adapting the techniques of the present paper directly
in the language of blossom trees. For quadrangulations and 
Eulerian triangulations,  both described by {\it binary} blossom
trees with buds and leaves (see Refs.[\xref\SCH,\xref\BMS] for details), this
involves weighting leaves, vertices and edges with suitable 
weights $\rho_i$ (per leaf at the right of a face at depth $i$
in the dual of the original graph), $g$ (per vertex), and $\alpha_j$
(per edge bordering a face of depth $j$ in the original graph).
Repeating this weighting procedure in more general cases of
blossom trees should give access to the local environment in 
so-called constellations, for instance two-constellations, i.e.
graphs with arbitrary even-sided faces, which generalize quadrangulations,
or three-constellations, which generalize Eulerian triangulations by allowing
white faces with valences multiple of $3$. Presumably, the 
generalization of the blossom-tree construction proposed in Ref.\CENSUS\
for arbitrary planar graphs should also allow to explore the local
environment of a vertex as well.

\bigskip
\noindent{\bf Acknowledgments:} 
We thank I. Kostov for pointing out Ref.\GKY. 
All authors acknowledge support by the EU network on ``Discrete Random Geometry",
grant HPRN-CT-1999-00161.

\listrefs
\end